\documentclass{elsart}

\usepackage{natbib}

 \usepackage{graphicx}
 \usepackage{epsfig}

\usepackage{amssymb}

\begin{document}

\begin{frontmatter}



\title{Adaptive Optics: Observations and Prospects for Studies of
  Active Galactic Nuclei}


\author{Richard Davies}

\ead{davies@mpe.mpg.de}

\address{Max-Planck-Institut f\"ur extraterrestrische Physik, 
  Garching, Germany}

\begin{abstract}
These lectures take a look at how observations with adaptive optics
(AO) are beginning to influence our understanding of active galactic
nuclei (AGN).
By focussing on a few specific topics, the aim is to highlight the
different ways in which enhanced spatial resolution from AO can aid
the scientific analysis of AGN data.
After presenting some background about how AO works, I will
describe a few recent observations made with AO of QSO host galaxies,
the Galactic Center, and nearby AGN, and show how they have
contributed to our knowledge of these enigmatic objects.
\end{abstract}


\end{frontmatter}

\section{Introduction}
\label{dav:intro}

Adaptive Optics (AO) serves two purposes.
It is an enabling technology for even more complex instrumentation
such as IR/optical interferometry.
But it is also an important technique in its own right.
At near-infrared wavelengths (1--5\,$\mu$m) it allows one to improve
the spatial resolution of ground-based telescopes by an order of
magnitude from 0.5--1$^{\prime\prime}$ to the diffraction limit, which
for an 8-m class telescope in the H-band (1.6\,$\mu$m) is about
40\,mas.
AO has applications in many areas of astronomy because it not only
increases a telescope's sensitivity for unresolved sources, but it
also reduces crowding problems in dense fields, and sharpens our view
of the morphologies and kinematics of extended objects.
In these lectures, the impact that adaptive optics is having on
studies of Active Galactic Nuclei (AGN) is discussed with particular
reference to three topics that span scales from $<1$\,pc in the
innermost region of our Galactic Center, through 10\,pc scales in
nearby AGN, to kpc scales in QSO host galaxies at high redshift.
By studying these different aspects and combining what we can learn
from different spatial scales, we can gain a more holistic view of the
structures comprising AGN and the physical processes governing how
they are fuelled.

\section{Adaptive Optics Overview}
\label{dav:ao}

It is only by understanding the basic ideas behind adaptive optics
that one can make the best use of it.
This section therefore begins by looking at the effect that
atmospheric turbulence has on a propagating wavefront.
By making the approximation that the turbulence occurs in a thin layer
at a fixed height, it addresses how adaptive optics can correct these
aberrations.
And it provides the tools needed to make simple quantitative statements
about the impact of adaptive optics on a series of observations.
Some ways in which the major limitations are being overcome are briefly
considered.

\subsection{Atmospheric Turbulence}
\label{dav:ao:atmos}

The atmosphere is a dynamically active structure which constantly has
energy injected into it, 
either through heating (e.g. direct sunlight or
re-radiation from the ground) or via convection currents and wind
shear.
These effects occur on scales of several tens of
metres (the outer scale, $L_0$), and the energy is gradually
transferred to smaller scales, until it is dissipated on scales of a
few millimetres (the inner scale, $l_0$).
This leads to local variations in the temperature and
density of the air, and hence in its refractive index $n$.
It can be quantified statistically using the structure
function which describes the difference in $n$ between a location
$r^\prime$ and another location separated from it by $r$.
For Kolmogorov statistics, the refractive index structure
function is defined in terms of an average  -- denoted
$\left<\right>$ -- over all possible $r^\prime$ and $r$.
It takes the form
\begin{equation}
D_n(r) \ = \ \left< [ n(r^\prime)-n(r^\prime+r) ]^2 \right>
     \ = \ C_n^2 r^{2/3}
\label{dav:ao:eq:risf}
\end{equation}
where $C_n^2$ is the refractive index structure constant.
It is notable that the $r^{2/3}$ here is the origin of all the 
$\frac{1}{3}$ powers in expressions associated with adaptive optics.
As it stands, this expression implies that the refractive indices at
two points decorrelate ever more as the separation between the points
increases.
But this is true only as long as $r<L_0$ (e.g. for 2-m or 4-m telescopes).
Instead, the decorrelation reaches a maximum at $r=L_0$, and the
effect of this can be measured even on an 8-m telescope.
As a result, the van Karman model is more often used. 
This is the same as the Kolmogorov model, but takes into account the
inner and outer scales.

The impact that $D_n(r)$ has on a wavefront propagating down through the
atmosphere is described using the phase structure function
\begin{equation}
D_\phi(r) \ = \ 2.91 
       \left( \frac{2\pi}{\lambda} \right)^2 
       \left( \sec{\zeta} \right)^{5/3}
       \int_{0}^{\infty} C_n^2(h) dh
\label{dav:ao:eq:dphi}
\end{equation}
a good derivation of which is given by \cite{rod81}.
In this equation, $\zeta$ is the angular zenith distance at which one
is looking, since the $C_n^2(h)$ profile is defined vertically through
the atmosphere. 
In the most simple approximation, one can think of the atmosphere
as a single thin `phase screen', which induces a relative (phase)
delay between two different points on a wavefront that depends only on
the lateral separation of those points.

An important quantity to introduce here is Fried's parameter $r_0$
\citep{fri65},
which describes the coherence length of the atmosphere and is defined as 
\begin{equation}
r_0 \ = \ 
    0.185 \ \lambda^{6/5}
    \left( \sec{\zeta} \right)^{3/5}
    \left( \int_0^\infty C_n^2(h) dh \right)^{-3/5}
\label{dav:ao:eq:r0}
\end{equation}
Put in simple terms, $r_0$ is the size of aperture (e.g. telescope
mirror) across which a wavefront can, for practical purposes, be
considered unperturbed.
In fact, the variance $\sigma^2$ of the resulting wavefront
aberrations over a circular aperture of diameter $D$ can be expressed
in terms of Fried's parameter as 
\begin{equation}
\sigma^2 \ = \ 1.030 \left( \frac{D}{r_0} \right)^{5/3}
\label{dav:ao:eq:sig}
\end{equation}
Thus `unperturbed' means having only 1\,rad$^2$ of variance.
The immediate conclusion one can draw from this equation is that
turbulence limits the resolution of a telescope to $\lambda/r_0$
instead of $\lambda/D$.
Other important parameters can be derived from $r_0$, two of which are
the coherence timescale $\tau_0$ and the isoplanatic
angle $\theta_0$.
However, note that in reality these parameters are much less well
correlated than implied here.

The coherence timescale can be thought of as the temporal equivalent
of $r_0$, and is related to the time taken for the wind to blow across
a cell of size $r_0$.
Specifically, it is the time over which the
variance of a wavefront changes by 1\,rad$^2$.
It is defined assuming Taylor's frozen flow
hypothesis which asserts that temporal changes in the turbulence along
a line of sight are due only to the lateral shift of an otherwise
fixed phase screen.
It is given as
\begin{equation}
\tau_0 \ = \ 0.314 r_0 / V_{wind} 
\ \ {\rm where} \ \ 
V_{wind} \ = \ \left[ \int_0^\infty C_n^2(h) v^{5/3}dh
               / \int_0^\infty C_n^2(h) dh \right]
\label{dav:ao:eq:tau0}
\end{equation}
where $v$ is the wind velocity as a function of height, and hence
$V_{wind}$ is a weighted mean wind velocity.

The isoplanatic angle is related to the angular size of $r_0$, which
depends on the weighted mean distance $H$ to the phase screen that we
are using to approximate the atmosphere.
Specifically, it is the angular distance between two lines of
sight which causes the wavefront variance to change by 1\,rad$^2$,
and is defined as
\begin{equation}
\theta_0 \ = \ 0.341 r_0 / H
\ \ {\rm where} \ \ 
H \ = \ \sec{\zeta} \left[ \int_0^\infty C_n^2(h) h^{5/3} dh 
                   / \int_0^\infty C_n^2(h) dh \right]^{3/5}
\label{dav:ao:eq:theta0}
\end{equation}
As we shall see later, the three parameters above form the basis for
estimating the wavefront variance due to the atmosphere:
$r_0$ and $\tau_0$ describe the ambient atmospheric conditions
and hence how well the AO system is likely to perform, while
$\theta_0$ describes how much the performance will degrade as
one moves off-axis from the guide star.
They all depend on $C_n^2(h)$, so that some knowledge of the 
turbulence profile (refractive index structure constant) through the
atmosphere is needed.
Various models for how $C_n^2(h)$ varies with height are
available, each typically including a strong ground layer and then
more moderate turbulence distributed in layers up to a height of
10--20\,km.
Direct measurements of $C_n^2(h)$ now typically form a crucial part of
the site characterisation for any modern telescope.

\subsection{Impact of a Perturbed Wavefront}
\label{dav:ao:impact}

\begin{figure}
\begin{center} 
\psfig{file=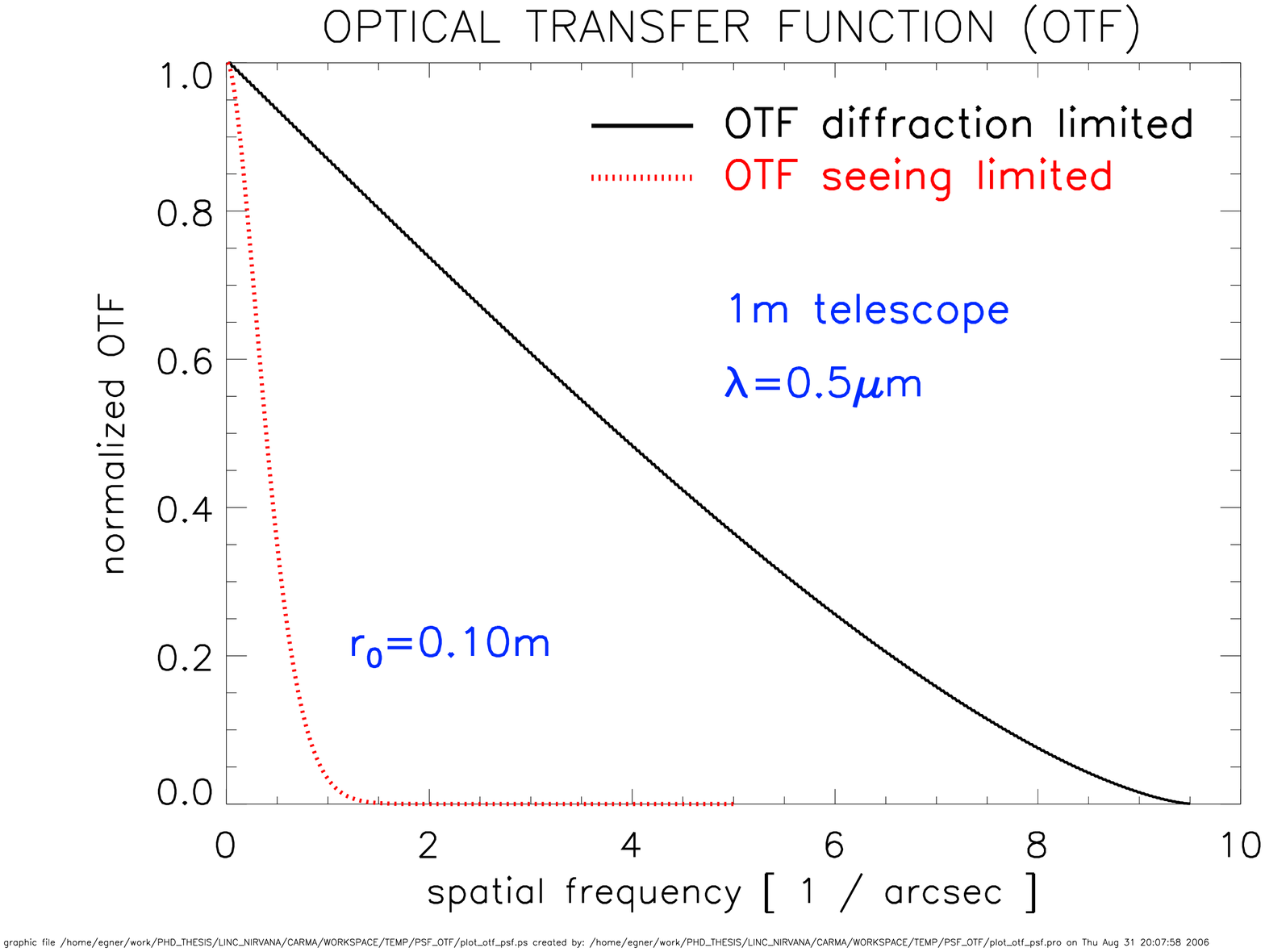,angle=90,width=6.5cm,clip=}
\psfig{file=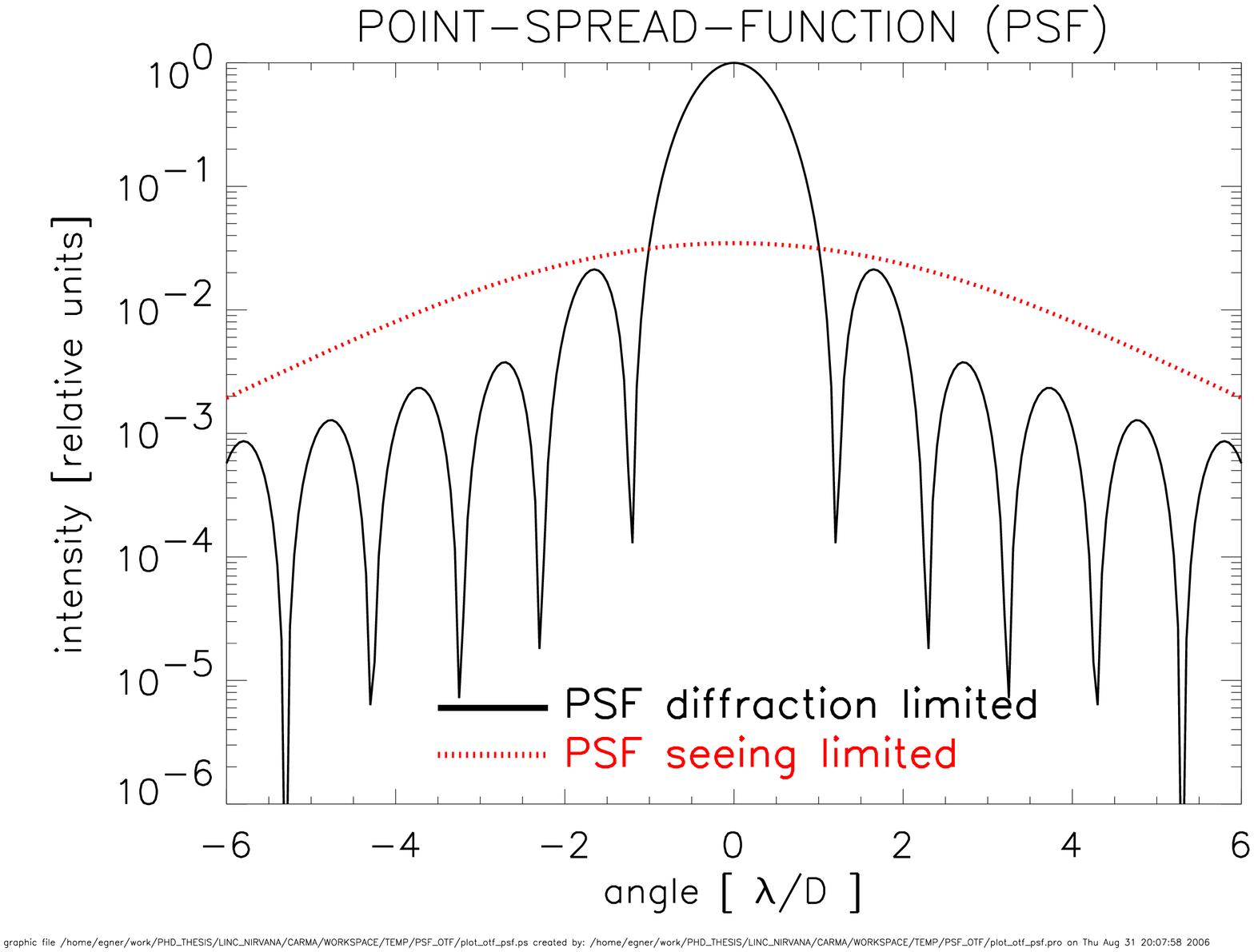,angle=90,width=6.5cm,clip=}
\end{center} 
\caption{Optical Transfer Function (left) and Point Spread Function
  (right) for perfect optics giving diffraction limited imaging, and
  also for a perturbed wavefront. The OTF and PSF are Fourier
  Transforms of each other; for a plane wavefront propagating through
  perfect optics, the OTF is the
  auto-correlation of the (entrance) pupil. From \cite{egn06},
  courtesy of S.~Egner.} 
\label{dav:ao:fig:otfpsf}

\end{figure}

After passing through the atmosphere, a wavefront that was originally
planar will be corrugated.
As such, the normal vectors at different points -- i.e. the direction
in which the photons are propagating -- are no longer parallel.
The direct result is that they can no longer be brought to focus at a
point, and instead form a blur.
This is quantified by the point spread function (PSF), which is the
image produced by an optical system of an unresolved source .
If a plane wavefront propagates through a circular entrance aperture
(pupil) and is focussed by perfect optics, the image will be an Airy
function which shows clear diffraction rings.
Mathematically, the pupil and PSF are related via the optical transfer
function (OTF) which describes how well different spatial frequencies
are transferred through the optical system.
The OTF is the autocorrelation of the pupil,
and the PSF is the Fourier transform of the OTF.
As Fig.~\ref{dav:ao:fig:otfpsf} shows, for a circular pupil, the OTF 
is very nearly a conical function.
A strongly perturbed wavefront will yield only a more diffuse PSF, which is
often characterised by a Gaussian function (albeit with broader wings).
In this case, the OTF is also approximately Gaussian, and shows that
high spatial frequencies are not transmitted through the imaging
system (i.e. the atmosphere).

\subsection{A Simple Adaptive Optics System}
\label{dav:ao:simpleao}

\begin{figure}

\hspace{1cm}
\includegraphics[width=7cm]{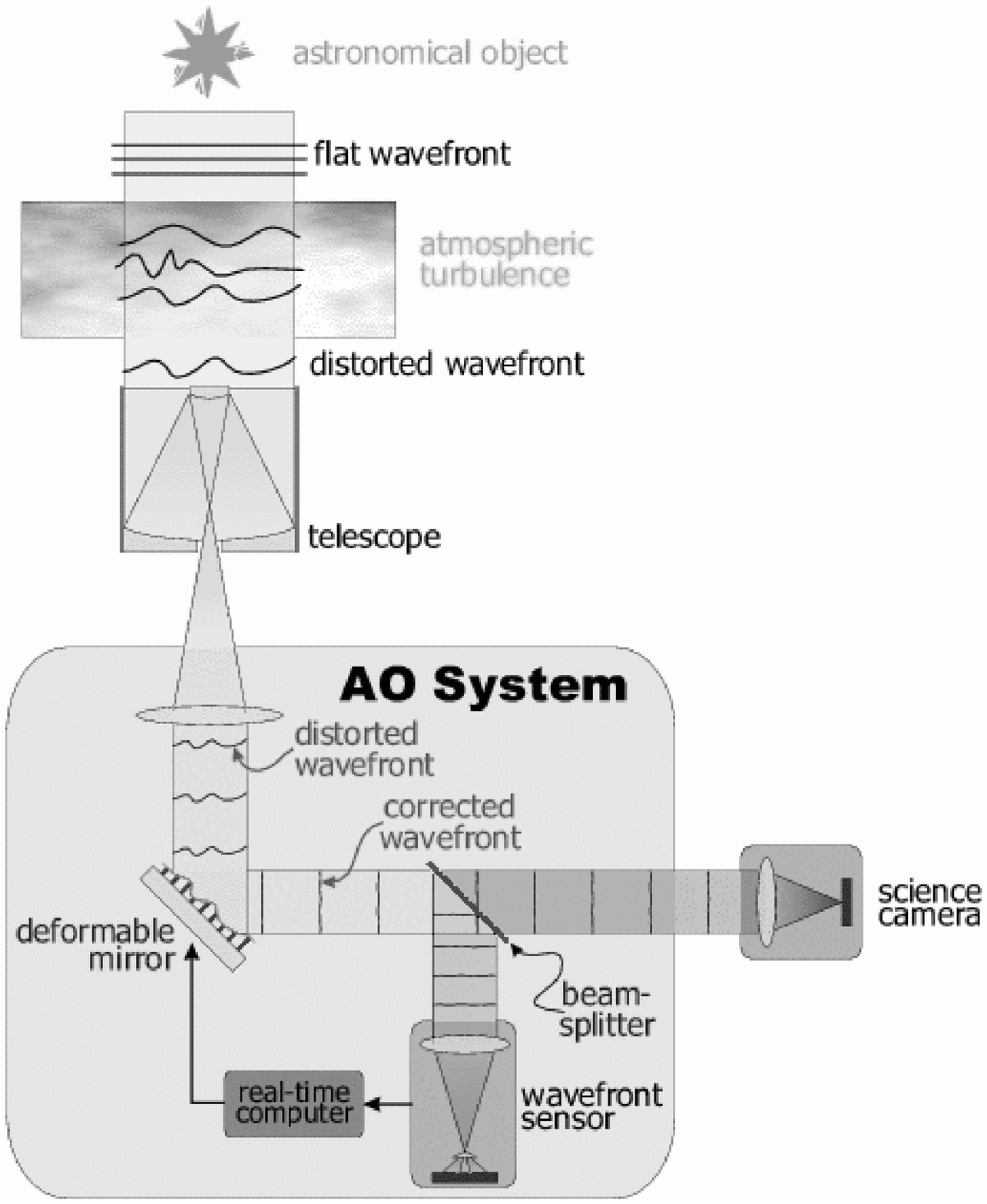}

\vspace{-7cm}
\hspace{8.5cm}
\includegraphics[width=4cm]{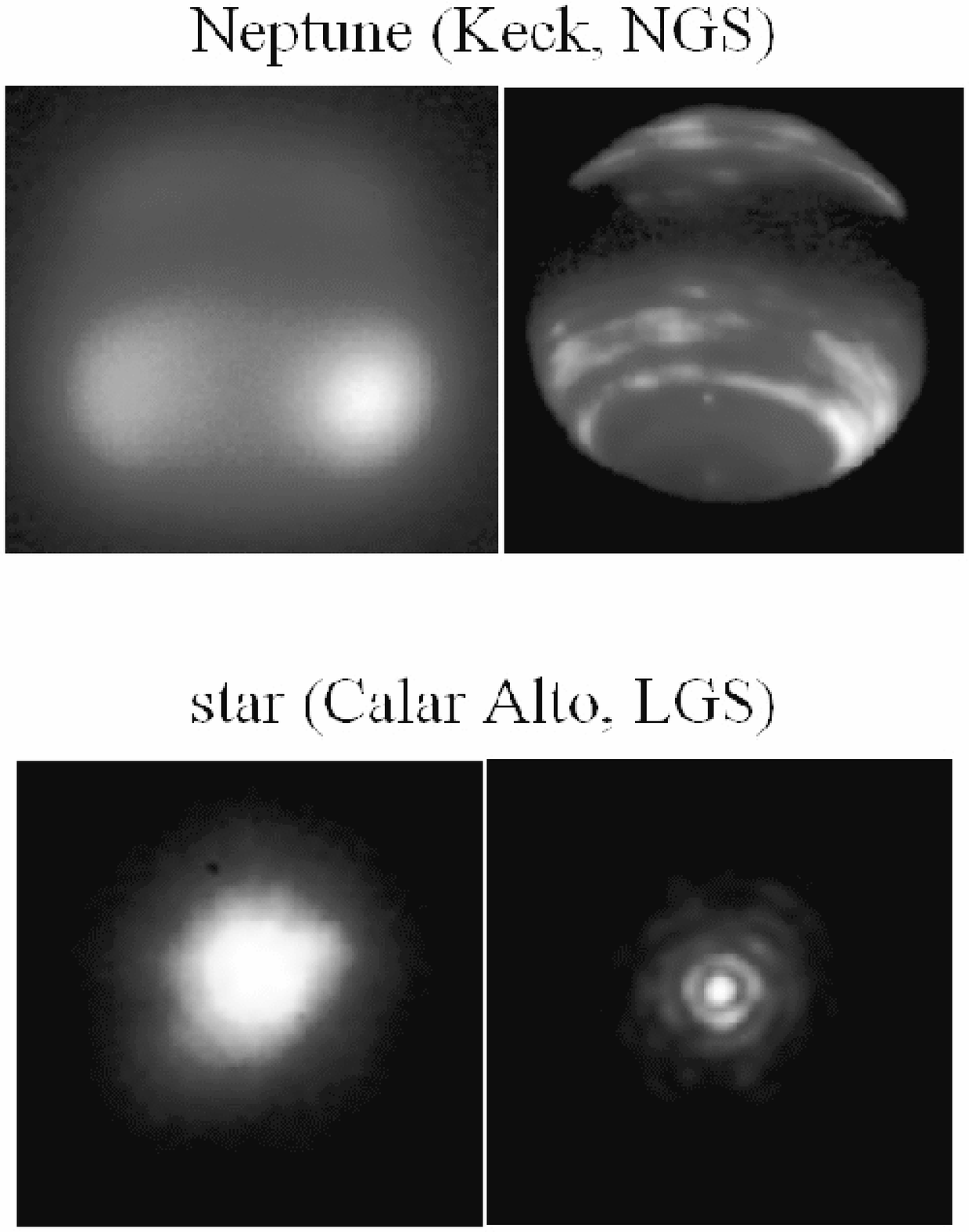}
\vspace{0.1cm}

\caption{Left: a cartoon showing the principle of a simple AO system;
 from \cite{egn06}, courtesy of S.~Egner.
Right: examples of seeing limited and corrected images.
Top: Neptune, taken on the Keck telescope using NGS, courtesy of C.~Max;
Bottom: a star corrected with the LGS-AO system ALFA, showing
 diffraction rings, adapted from \cite{dav00}.}
\label{dav:ao:fig:simpleao}
\end{figure}

To flatten a wavefront that has been corrupted by the atmosphere, an
AO system must include a wavefront sensor (WFS) to measure the shape
of the wavefront and a deformable mirror (DM) with which to apply the
corrections (although for practical purposes the global slope, or
tip-tilt, is usually corrected by a separate flat mirror).
What happens can be seen in Fig.~\ref{dav:ao:fig:simpleao}. 
The perturbed wavefront is reflected from the initially flat DM 
into the WFS. 
Here, the shape of the wavefront is measured, and a real-time computer
turns these measurements into a set of commands (i.e. voltages) which
are sent to the DM.
The DM can now correct the shape of the wavefront to be what it was
a short time previously (at least to the limits of the WFS
sensitivity and mechanical ability of the DM).
Thus, what remains is only a small residual wavefront aberration due to
changes since the previous measurement was made.
It is this residual aberration that is now measured by the WFS, and a
correction is made to update the DM commands.
This update cycle of only the residual wavefront error
is known as `closed loop' operation.

The fact that the wavefront is sensed at optical wavelengths, while the
science channel is infrared implicitly assumes that the same
correction is valid for both wavelengths.
This is true because
the wavefront aberrations are simply shifts in longitudinal position
of the wavefront at different places.
And a shift (i.e. delay) of, say, 100\,nm is the same whether it
applies to infrared or optical photons.
The wavelength dependence arises from the impact of this delay:
100\,nm is $\lambda/5$ in the optical, but it is
only $\lambda/20$ in the near infrared and hence has less effect there.
We have already seen this in
Fried's parameter, Eq.~\ref{dav:ao:eq:r0} showing that 
$r_0 \propto \lambda^{6/5}$. 
Thus, at longer wavelengths, the coherence length and timescales are
longer.
As such, the image quality is not degraded so much, and adaptive
optics can work better.

\begin{figure}
\begin{center} 
\psfig{file=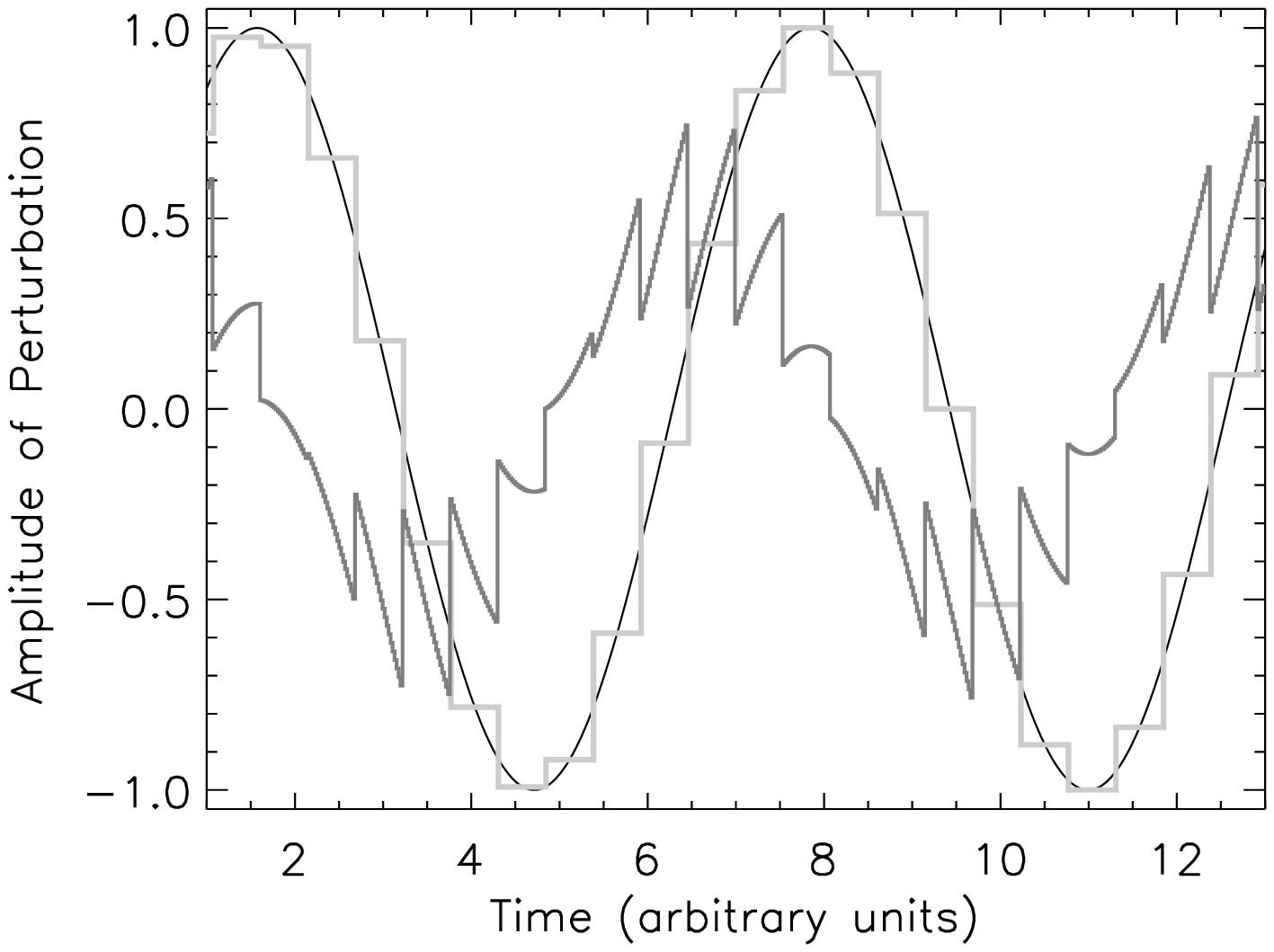,width=7.0cm}
\psfig{file=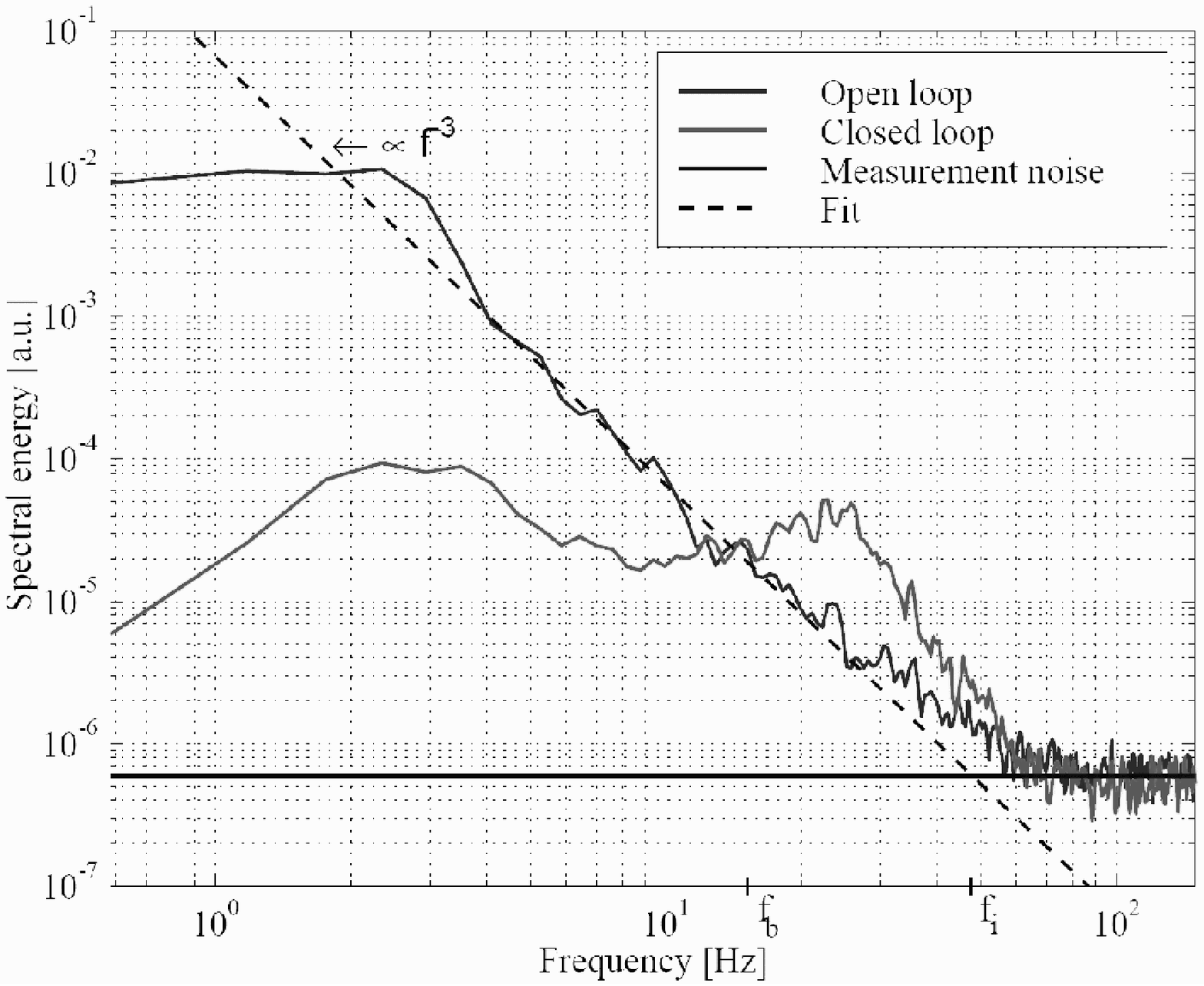,width=6.4cm}
\end{center} 
\caption{Left: Illustration showing that if a perturbation (black
  line) is
  sampled at 12 times the speed at which it changes (light-grey
  line), the strength of the residual (dark grey line) is reduced by
  only a factor 2 
  but also exhibits higher frequency perturbations.
Right: Data measured with the ALFA AO system showing that in close
  loop, the low frequency perturbations are corrected, but extra
  energy is put into the higher frequencies.
The AO system was running with a frame rate of 300\,Hz but the closed
  loop bandwidth $f_b$ is only 15\,Hz. The noise floor $f_i$ is
  reached at about 50\,Hz. From \cite{kas00}, courtesy of M.~Kasper.}
\label{dav:ao:fig:speed}
\end{figure}

Perhaps one surprising feature of AO is the speed (frame rate)
at which the wavefront must be measured in order to correct it, as
demonstrated in Fig.~\ref{dav:ao:fig:speed}.
Typically, the bandwidth (i.e. the frequency below which aberrations
are well corrected) is about 1/10 the frame rate.
This becomes apparent when comparing $\tau_0$ to the Greenwood
frequency $f_G$, which 
is defined from the engineering perspective as the closed loop
bandwidth at which the residual wavefront variance is 1\,rad$^2$.
One finds that $1/\tau_0 = 7.5f_G$, a difference that is due to the
inevitable time delay that arises because a 
correction can only be applied after it has been measured.
If the AO system has a frame rate of 100\,Hz, the shape of the DM can
only be updated at least 10\,ms after the measurement of the wavefront
residual has started.
It means that algorithms which can predict what shape the
wavefront will have even a short time in the future, could
yield a significant improvement in AO performance.
This issue, however, is rather complex because the atmosphere
consists of multiple layers moving in different directions, and there
is no easy way to make a sensible prediction unless the different
layers can be sensed separately.
It is therefore reassuring that some progress is being made in this
direction using Fourier techniques \citep{poy07}.

\subsection{Wavefront Sensors}
\label{dav:ao:wfs}

There are a number of ways of measuring the shape of the
wavefront, 
the two most commonly employed of which are the Shack-Hartmann sensor
(developed by \citealt{sha71}, based on a device originally devised by
\citealt{har00}) 
and the curvature sensor \citep{rod88}.

A Shack Hartmann sensor uses an array of lenslets to split the pupil
into a series of sub-apertures, and is easy to extend to very high
order.
However, the geometry cannot be adjusted to
match the ambient seeing conditions,
and the resolution is limited by the sub-aperture size rather than the
full pupil.
It works by creating an image of a star through each sub-aperture,
the position of which depends on the local wavefront slope.
The combination of all of these local slopes yields the global first
derivative of the wavefront.
Such systems are typically used with a DM consisting of a series of
individual piezo actuators, across which is glued a thin flexible
mirror.
The actuator lengths change as voltages are applied,
generating local gradients across the mirror surface.

A curvature sensor measures the second derivative of the wavefront. 
It does this by comparing slightly pre- and post-focal images of the
star, and the difference in intensity at each point yields the local
wavefront curvature. 
This type of sensor is usually used with a bi-morph mirror, consisting
of a 2-layer piezo ceramic, one surface of which is silvered.
It bends when a voltage is applied at a certain point,
naturally reproducing the measured curvature.
Although it is harder to make high order systems this way, there are a
number of advantages:
the sensitivity can be adjusted at any time by modifying how afocal
the two reference images are; 
and the aperture is not divided, so one can benefit from the higher
resolution achieved in closed loop.


\subsection{Residual Wavefront Variance}
\label{dav:ao:res}

In Section~\ref{dav:ao:atmos} we saw how $r_0$, $\tau_0$, and
$\theta_0$ could be estimated 
from a measurement of $C_n^2(h)$ in the atmosphere.
These characterise the spatial and temporal distortions of the
incoming wavefronts, which are then measured and corrected by an AO
system which itself has some spatial and temporal limitations.
Using these, one can estimate how the corrected PSF
should look -- and quantify it in terms of the Strehl, the ratio of
the actual peak intensity to that which would be achieved 
with a perfect optical system.

Spatial wavefront distortions are usually characterised in terms
of orthonormal basis functions, the simplest of which for a circular
aperture are the Zernike modes.
The first few low order Zernike terms describe the well known effects of
piston, tip-tilt, astigmatism, focus, coma, and trefoil.
However, since telescope apertures are annular, and annular Zernike modes
are not orthonormal, in practice the more
complex Karhunen-Lo\`eve modes are used.
Only a finite number of modes can be corrected
(no more than about 2/3 the number of measurements made), leaving a
residual wavefront variance 
\begin{equation}
\sigma^2_{fit} \ \sim \ 0.2944j^{-\sqrt{3}/2} (D/r_0)^{5/3}
\label{dav:ao:eq:sfit}
\end{equation}
where $j$ is the number of (Zernike) modes corrected.
The expression is valid for large $j$;
variances for small $j$ are given in \cite{bec93}.
The residual variance due to the limited bandwidth of the AO system
can be estimated as 
\begin{equation}
\sigma^2_{lag} \ = \ (\tau/\tau_0)^{5/3}
\label{dav:ao:eq:slag}
\end{equation}
where $\tau$ is the time lag from the start of a measurement with
the WFS to the time at which it is applied to the DM.
And finally, the residual variance which occurs if the science
target is a finite angular distance $\theta$ from the guide star is
\begin{equation}
\sigma^2_{angle} \ = \ (\theta/\theta_0)^{5/3}
\label{dav:ao:eq:sang}
\end{equation}
There are inevitably other constraints on the performance of any AO
system, such as the photon and read noise in the measurement.
These depend on the design and hardware of each specific
system and are discussed more elsewhere \citep[e.g.][]{har98}.
Although beyond the scope of this brief overview, they need to be
included in the estimate of the total variance which is then used
to estimate the Strehl $S$ with the approximation
\begin{equation}
S \ \sim \ e^{-\sigma^2_{total}}
\ \ {\rm where} \ \ 
\sigma^2_{total} \ = \ \sigma^2_{fit} + 
                       \sigma^2_{lag} + 
                       \sigma^2_{angle} + 
                       \sigma^2_{noise} + ...
\label{dav:ao:eq:strehl}
\end{equation}
Some typical values for the K-band (2.2\,$\mu$m) Strehl on an
8-m telescope are:
1--2\% for the seeing limit; 
20\% for reasonable performance; 
50\% with on-axis bright stars; 
and $>90\%$ is the aim for the generation of extreme AO systems
currently being designed.

\subsection{Sodium and Rayleigh Laser Guide Stars}
\label{dav:ao:lgs}

The major limitation with natural guide star (NGS) AO is that for good
performance one needs a star brighter than $V\sim13$\,mag -- and the
probability of finding one close to any particular science target is
very low.
As a result, the isoplanatic residual variance $\sigma^2_{angle}$
often dominates the error budget.
To keep this term under control, the sky coverage for NGS-AO is
limited to at most a few percent, and far less at the galactic pole.
The alternative is to generate an artifical star, which one can point
anywhere in the sky. 
This can be done by projecting a high quality and high power laser
beam into the sky and measuring the back scattered light.
Rayleigh lasers (usually green or UV) are pulsed and one can gate the
detector so it sees each pulse of light scattered by the air at a
specific altitude, usually in the range 10--30\,km.
Sodium line lasers (orange) are designed to make the layer of neutral
sodium atoms at a height of $\sim$90\,km fluoresce.
Although using a laser to create an artifical star does to a large
extent solve the isoplanatic issue, it also brings its own problems.

A Rayleigh laser system is relatively straightforward to construct.
However, for 8-m class telescopes there is a major limitation known as
the cone effect (or focal anisoplanatism).
Because the LGS is at a finite height, its light only samples a
conical volume along the line of sight, whereas light from a distant
astronomical object traverses a cylindrical region through the
atmosphere.
This means that much of the turbulence is not sensed and so cannot be
corrected. 
To avoid this issue, one can either use multiple guide stars (see
Section~\ref{dav:ao:mcao}) or create a guide star much higher up in
the atmosphere. 

For this reason, sodium line lasers are a popular option.
However, a major issue is the technology needed to build
them. 
As yet there is no robust turn-key laser that
can produce of order 10\,W at 589\,nm.
In addition, the mesospheric sodium layer is rather thick (typically
10\,km) and so any spot seen off-axis will be elongated --
a critical issue even on 8-m class telescopes if the
laser is not projected from behind the secondary mirror.
Furthermore, the sodium layer is a dynamic entity, and the number of
atoms is very variable, impacting the brightness of the spot.
But also the mean height can vary rapidly, dominating over the change
in height due to tracking across the sky.
If one does not correct for both of these, one can accumulate
significant focus residuals.
Finally, as for Rayleigh lasers, one still needs a NGS to correct for tip-tilt
since this cannot be measured with an LGS (the laser accumulates an
unknown tip-tilt as it propagates upward through the atmosphere).
While a tip-tilt NGS can be fainter and further from the science object than
one used for high order AO correction, it is still a significant
constraint.
As yet, none of the solutions proposed have been satisfactory.
As an alternative, it has been suggested to dispense with tip-tilt if
one does not need to reach the diffraction limit \citep{dav08}.

\subsection{Multi-Conjugate Adaptive Optics}
\label{dav:ao:mcao}

The residual variance due to isoplanatism not only limits the sky
coverage -- which led to the development of LGS -- but it limits the
field of view that can be corrected.
At Paranal, the typical K-band isoplanatic angle is only
12.7$^{\prime\prime}$ \citep{cre06}, so even at this distance from
the guide star the Strehl is much reduced.
This obstacle can be surmounted using multi-conjugate adaptive optics
(MCAO).
Here one uses multiple laser (or natural) guide stars to fully sample
the turbulence.
With a tomographic reconstruction, one can then approximate the
atmosphere as several phase screens at different heights.
It is then possible to use two or more DMs, each conjugated to a
different layer, to perform a full correction on the wavefront that
works even off-axis. 
Essentially what is happening is that off-axis rays reflect off
the DMs in different places to an on-axis ray, and hence receive 
different total corrections that are appropriate for their path
through atmosphere.
Over recent years, there has been a considerable effort in this
direction, leading to two concepts for MCAO systems.
The first uses classical tomography: the wavefront errors for each
guide star are sensed separately; and the measurements are combined in a
computer which calculates how much phase delay occured at each point
in each layer.
In layer oriented MCAO \citep{rag00}, the light from the guide stars
is combined optically on the detectors, and there is one detector
associated with one DM for each layer that is corrected.
For this design, each layer/detector/DM combination is an independent
AO loop, and hence the whole system is computationally much simpler.

Although a proto-type MCAO system for solar astronomy was tested in
the last few years on a small telescope, 
the first one for an 8-m class telescope became operational only recently
\citep[MAD, ][]{mar07}.
During the next few years, several others should become operational on
Gemini South, the Very Large Telescope, and the Large Binocular
Telescope.

\subsection{Realistic Expectations}
\label{dav:ao:expect}

\begin{figure}
\begin{center} 
\psfig{file=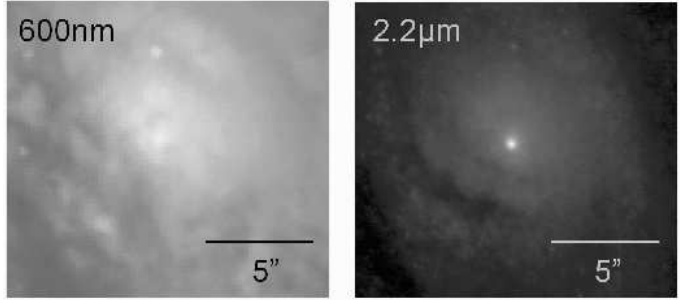,width=9cm}
\end{center}
\caption{Optical and near infrared views of the
  nucleus of the Circinus galaxy, on a logarithmic scaling.
While the nucleus is faint and fuzzy at 600\,nm, there is a bright
  compact source suitable for an infrared WFS at 2.2\,$\mu$m.
Optical image courtesy of the {\em HST} archive; infrared image from
  \cite{pri04}.}
\label{dav:ao:fig:circinus}
\end{figure}

The discussion so far has assumed that the AO guide star is just that:
a star.
So what should we expect for AGN, which in general are not
particularly bright, are often fuzzy at optical wavelengths, and have
a relatively bright background due to the host galaxy?
One of the clearest examples of this problem is the Circinus Galaxy.
Fig.~\ref{dav:ao:fig:circinus} shows that at optical wavelengths there
is almost nothing to use as a wavefront 
reference, and so AO cannot perform well.
It is hard to predict what performance one should
expect using an AGN itself as the wavefront reference.
In such cases, one can often do much better than the seeing limit, but
one should certainly not expect perfect performance every time.
Fortunately, Circinus does have a bright compact near infrared source
suitable for AO guiding if one has an infrared wavefront sensor like
that in NACO on the VLT.
Alternatively -- and more commonly -- one will have to use a laser
guide star.
But one should be aware that with current LGS facilities, one can get
anything from 0.1$^{\prime\prime}$ resolution to $\sim$30\% Strehl in the
K-band depending on the ambient conditions.
It is worth remembering that LGS-AO is still in its infancy, and there
are likely to be many improvements in future generations of LGS-AO
systems that should lead to greater efficiency and better performance
for more of the time.

\section{Quasar Host Galaxies}
\label{dav:qso}

Armed with some knowledge about adaptive optics, we can explore where
it can make the biggest impact in studies of AGN. 
We begin with QSOs, where the aim has been to detect and measure the
properties of their host galaxies. 
We highlight cases where AO has started making an impact on this topic, and
discuss the implications of this work in the context of
where the most powerful QSOs reside, and of the possible evolutionary
scenario linking them to ultraluminous galaxies (ULIRGs). 

\subsection{What can we learn from QSO Host Galaxies?}
\label{dav:qso:what}

Since its discovery, the relation between the mass $M_{\rm BH}$ of the
black hole in the
center of a stellar spheroid and the velocity dispersion $\sigma_*$ of
that spheroid \citep{fer00,geb00} -- which superseded similar ones
using the luminosity or mass of the spheroid \citep{kor95,mag98} --
has become a cornerstone for cosmological models.
It is generally accepted that the $M_{\rm BH}-\sigma_*$ relation should be
valid for all spheroids irrespective of mass-scale, environment,
and evolutionary history -- for example, whether the black hole is
quiescent or active, whether the spheroid is embedded in a gaseous
disk or not, or whether it is a giant elliptical or a globular cluster.
This interpretation is bolstered by evidence that moderate mass
systems such as Omega Centauri host intermediate mass black
holes \citep{noy08}.
And it only makes sense if black holes grow at the same rate as the
bulges in which they reside.
It is this pretext that currently
forms the basis for our understanding of galaxy
evolution and black hole growth.

Over the last decade there has also been an intense effort to derive
the evolution with cosmic time of both the star formation rate density
and the number density of luminous QSOs.
Remarkably, these both follow qualitatively similar trends and both
peak around $z \sim 2$--3, at least indicative that stars and AGN do
indeed evolve and grow in parallel.
However, as always there are a few glaring exceptions, of which
perhaps the most spectacular is the $z=6.4$ QSO J1148+5251.
Continuum and emission line scaling relations yield
$M_{BH} = (1-5) \times 10^9$\,M$_\odot$ \citep{fan03}.
But the fact that such a massive black hole can exist so early in the
universe implies sustained and very rapid growth.
To add to the puzzle, interferometric
observations of the molecular gas in this object 
were able to resolve the CO(3-2) line spectrally and
spatially \citep{wal04} .
They revealed two peaks separated by 1.7\,kpc.
Assuming that these are gravitationally bound with a
rotational velocity corresponding to the 280\,km\,s$^{-1}$ FWHM of the
line, the dynamical mass of the system is $\sim5\times10^{10}$\,M$_\odot$.
This is more than an order of magnitude less than the
$\sim10^{12}$\,M$_\odot$ implied by
the $M_{\rm BH}-\sigma_*$ relation.
This result suggests that at early times, black holes may form before
stellar bulges are assembled.

It is to understand more about the $M_{\rm BH}-\sigma_*$ relation and
its relevance and implications for cosmic evolution that we need to
study QSOs and their host galaxies.
One way to do this is to look at local relic populations and infer
what must have happened to create them; or we can look directly at 
high redshift QSOs where the action is actually taking place.

\subsection{How can we study QSOs and their Hosts?} 
\label{dav:qso:how}

At low redshift one can use reverberation mapping
(RM) to derive $M_{\rm BH}$ by equating it with the  
virial product of the width of an emission line from the broad line
region (BLR) and the time delay with which its flux follows changes in
the UV continuum \citep{pet01}.
Secondary relations then make use of the apparent scaling between the
size of the BLR and the optical continuum luminosity
\citep{kas05} to extend the RM technique to higher redshift,
using the virial product of an emission line FWHM and the BLR size.
This is simply asserting that $M \propto \sigma^2 R$ which, when
calibrated, becomes \citep{ves04}
\begin{equation}
\log{ \frac{M_{BH}}{M_\odot}} 
   \ = \ 6.7 + 0.7 \log{\left(
   \left[ \frac{{\rm FWHM(H\beta)}}{1000\,km\,s^{-1}} \right]^2
   \left[  \frac{\lambda L_\lambda (5100{\rm \AA})}
              {10^{44}\,erg\,s^{-1}} \right]^{0.7} \right)}
\label{dav:qso:eq:ves}
\end{equation}
The Eddington rate $L_{bol}/L_{Edd}$ can then be found from the QSO
luminosity since
\begin{equation}
L_{Edd} / L_\odot \ = \ 3.3 \times 10^4 M_{BH} / M_\odot
\label{dav:qso:eq:edd}
\end{equation}

\begin{figure}
\begin{center}
\psfig{file=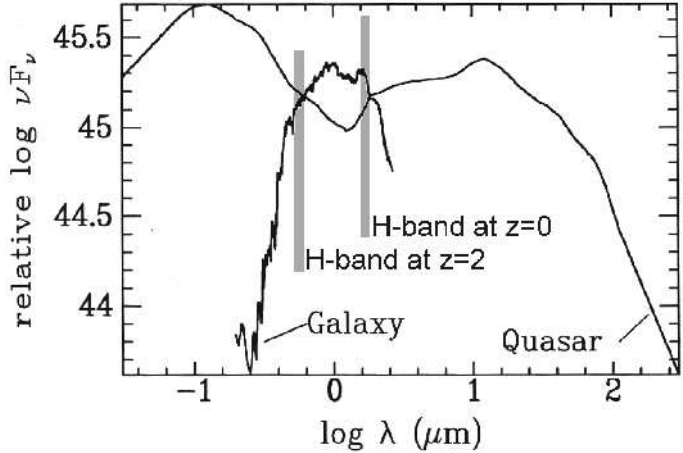,width=9cm}
\end{center}
\caption{Spectral energy distributions of a typical QSO and galaxy
  (arbitrarily scaled). The rest-frame location of the H-band for
  $z=0$ and $z=2$ spans the region where the host galaxy
  contrast is best. 
Adapted from \cite{mcl95}.}
\label{dav:qso:fig:contrast}
\end{figure}

One then just needs to compare these to the properties of the host
galaxy -- which is where AO is important because spatial resolution is
crucial to the success of this venture.
From an astrophysical perspective, Fig.~\ref{dav:qso:fig:contrast}
shows that the near infrared is the best wavelength regime for
studying QSO host galaxies.
The QSO spectrum is dominated at short wavelengths by thermal emission
from an accretion disk at $\sim10^5$\,K, and at long wavelengths by
reprocessed emission from dust at $10^2$--$10^3$\,K, with a minimum in
between at near infrared wavelengths.
It is at such wavelengths that emission from the host galaxy, which is
dominated by star light, reaches its maximum.
An additional advantage of the near infrared regime is that one
suffers less from the effects of extinction than optical; 
and at high redshift there is less bias towards only very recent star
formation which is traced by the rest-frame UV continuum.
In principle one could use HST/NICMOS, but AO on ground-based 8-m
class telescopes can provide superior results due to the better
spatial resolution, as well as the longer integrations and larger
surveys that are needed \citep{hut06}.

\subsection{QSOs and their Hosts at Low Redshift}
\label{dav:qso:lowz}

The largest adaptive optics survey of QSO host galaxies to date was performed
by \cite{guy06} using the Gemini North and Subaru telescopes.
They observed 32 nearby ($z<0.3$) QSOs, mostly in the H-band
(1.6\,$\mu$m), with a dynamic range of $>10^4$ so that the host
galaxies could be traced out to several arcseconds.
PSF reference stars were observed for every target, using the same
instrumental set-up.
This is an important step, because to model the host galaxy, the
bright QSO point source -- including its extended wings -- must be
subtracted reasonably accurately.
The scheme is outlined in some detail below, to demonstrate that this
process is far from trivial, and requires subjective human input as
well as an automated fitting routine.
The authors modelled each host as sum of an exponential (appropriate
for disks) or a de~Vaucouleurs $r^{1/4}$ law (appropriate for
spheroids).
These can be quantified with a S\'ersic profile using $n=1$ or
$n=4$ respectively as
\begin{equation}
I(r) \ = \ I_e \exp(-b_n [(r/R_e)^{1/n} - 1] )
\label{dav:qso:eq:sersic}
\end{equation}
where $R_e$ is the effective radius and $b_n$ is a constant that
varies with $n$ to ensure that $R_e$ encloses half the light.
Each profile requires 4 parameters -- brightness, size, orientation, and
axis ratio -- so that in Eq.~\ref{dav:qso:eq:sersic} they replaced $r$
with $f(r)$ for which
\begin{equation}
f(r) \ = \ r \sqrt{1 + \sin^2{(\theta-PA)} 
             \left[ \frac{1}{(1-e)^2} - 1 \right]}
\label{dav:qso:eq:fr}
\end{equation}
where $PA$ is the orientation and $e$ the elongation.
In addition, the scale of the central point source is unknown, giving
in total 9 free parameters.
These were fit by minimising the residuals, which were quantified as
\begin{equation}
\chi^2 \ = \ \sum \left\{ \left[ image - (pointsource + host)
             \otimes PSF \right] / \sigma^2 \right\}
\label{dav:qso:eq:min}
\end{equation}
This is an idealised scheme and cannot account for tidal tails or
spiral arms.
The residuals were therefore classified by eye, and a cross-check made
to ensure that, for example, spiral arms only occur in disky hosts.
In addition, the results for different bands were cross-checked manually.

\begin{figure}
\begin{center}
  \psfig{file=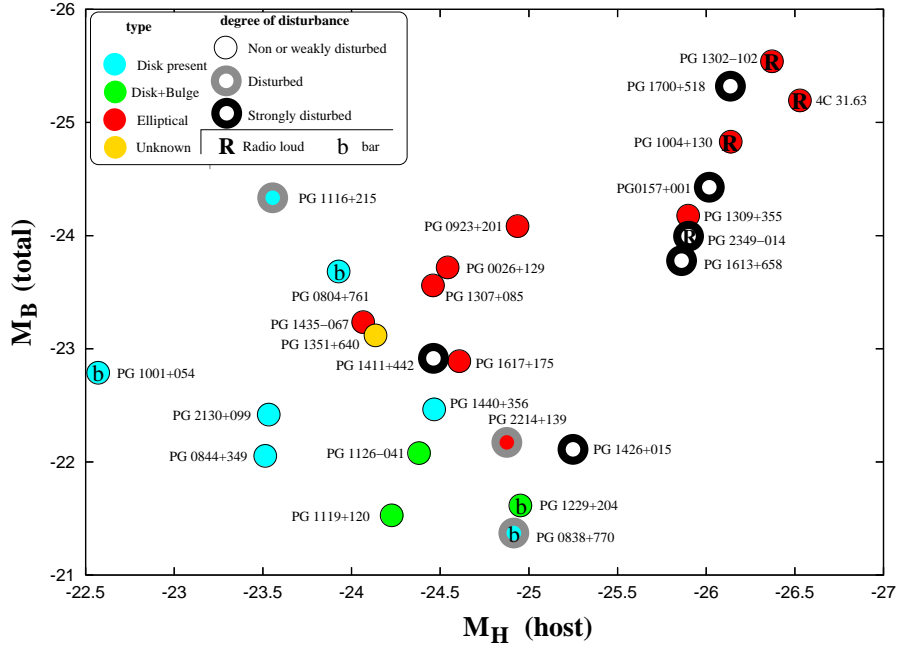,width=12cm}
\end{center}
\caption{Absolute B-mag of the QSOs plotted against absolute H-mag of
  the hosts for 27 objects. Different symbols and shades denote the
  morpholgoy of 
  the host (disk or bulge or both) and the strength of any
  disturbances due to an interaction. From \cite{guy06}, courtesy of
  O.~Guyon.}
\label{dav:qso:fig:guyon}
\end{figure}

Fig.~\ref{dav:qso:fig:guyon} neatly summarises the three important
results from this work.
\begin{enumerate}
\item
Lower luminosity hosts are mostly disks and
high luminosity hosts are mostly ellipticals, but with a significant
fraction of disturbed morphologies.
This suggests that luminous hosts are the result of galaxy
interactions or merger events. 
\item
Disky hosts do not harbour really luminous QSOs.
These are found in strongly disturbed or elliptical hosts.
This suggests that luminous QSOs only occur in really major mergers
which destroy the disk.
\item
Radio loud QSOs (which comprise $\sim$15\% of the total, a typical
fraction) are only found in the most luminous hosts.
With the $M_{BH}-\sigma_*$ relation (and bearing in mind that the
fundamental plane of early type galaxies implies $\sigma$ is related to
size and luminosity), this then suggests that the ratio of
radio-quiet to radio-loud QSOs correlates with black hole size or
perhaps black hole spin \citep[see also][]{sik07}.
\end{enumerate}
One further result that \cite{guy06} found concerned the amount of
dust in the QSOs.
This was quantified as the ratio between the reprocessed infrared luminosity
$L_{IR}$ and the direct optical luminosity $L_{BB}$ (see
Fig.~\ref{dav:qso:fig:contrast}).
They found that elliptical hosts, which harboured the most luminous
QSOs, also had the
lowest ratios of $L_{IR}/L_{BB}$ and hence were less obscured by dust.
On the other hand, disky hosts typically had high ratios and hence
more dust.

Taken together, these results strongly support current ideas about the
evolutionary origin of QSOs, a 
scenario that was first put forward by \cite{san88}
and has been supported observationally and theoretically by many others.
It proposes that ULIRGs are formed by the merger
of gas rich spiral galaxies.
As this happens, gas falls to the centre where it can fuel both a
starburst and an obscured AGN.
Gradually the dust is cleared by radiation pressure and supernova, to
reveal an optically bright QSO.
When the fuel supply runs out, the QSO fades, leaving an old
elliptical remnant.
However, as we shall see later, this may not be the full story.

\subsection{QSOs and their Hosts at High Redshift}
\label{dav:qso:highz}

At high redshift, this game is much more challenging, primarily
because the QSO is often more luminous, but also because $(1+z)^4$
surface dimming makes the host much fainter.
And because the angular size of the host is smaller at high redshift
(at $z>2$, 1$^{\prime\prime} \sim 8$\,kpc), knowing the PSF becomes
even more crucial.
Thus, even detecting the host can often be difficult.
For example, \cite{cro04} detected only 1 of 9
hosts at $z\sim2$ using AO on Gemini North, and \cite{fal05} detected
only 1 of 3 hosts at $z\sim2.5$ with AO on the VLT.
Fortunately, even non-detections can yield useful luminosity limits.
On the other hand, with an early AO system on a smaller telescope,
\cite{kuh05} detected all 3 hosts they observed.
Although the host luminosities were consistent with the QSO/host
luminosity relation, they were remarkably bright for their sizes.
This implied a lower mass-to-light ratio, and hence a
younger stellar population.
The authors showed that if the age was 100\,Myr at $z\sim2.2$, then
the hosts would fade onto the usual magnitude/size relation by
$z\sim0$.

These results all have one aspect in common: they reduced the problem to
estimating the host size and magnitude by assuming an
elliptical profile.
However, this assumption is not necessarily well justified.
Using AO on the CFHT to observe a sample of nearby
($z<0.6$) QSOs, \cite{mar01} showed that there can be a bias in
modelling the radial profile.
They found that while elliptical hosts were always fit as ellipticals,
disky hosts were also sometimes fit as ellipticals if the QSO was
bright.
\cite{hut06} went so far as to state that ``claims that all hosts
at $z>2$ are elliptical are suspicious.''
He has good reasons, and has shown that observations are strongly
affected by dynamic range:
bright inner bulges can be relatively easily seen and resolved at
0.1$^{\prime\prime}$ scales, while fainter extended parts which may
show signs of merging are much harder to detect.
This concern was also shared by \cite{can06} who reported deep
HST/ACS observations of a QSO at $z\sim0.2$ which has an elliptical
host. 
Their data revealed faint shell structures, indicative of a merger event
$\sim$1\,Gyr in the past which would have triggered a burst of star
formation.

These results highlight an on-going discussion about the stellar
populations of QSO host galaxies: are they $\sim$10\,Gyr old as one
would expect for elliptical galaxies, or is there evidence for more
recent star formation?

\subsection{Are there Young Stars in the Centres of QSO Hosts?}
\label{dav:qso:youngstars}

This debate is covered in a recent review of QSOs and their host
galaxies by \cite{lac06}, but the two
perspectives can be summarised as follows.
In a carefully selected sample of luminous QSOs, \cite{dun03} found
that all the hosts were quiescent ellipticals, perhaps with old
populations. 
Supporting this, spectroscopic work by \cite{nol01} found evolved
stellar populations with ages $\sim$10\,Gyr and only a small amount of
recent star formation.
In contrast, \cite{kau03} found that the colours of the hosts of
luminous QSOs were bluer than expected.
This and other spectroscopic indicators implied significant star
formation in the last 1--2\,Gyr.
\cite{can06} re-examined the Dunlop sample, finding
that in 13 of the 14 for which they obtained spectra, there was a
population of stars making up at least 10\% of the host's mass that
had formed in the last 0.6--2.2\,Gyr.
But it is hard to understand how any event -- even a major one
such as a merger -- 1\,Gyr or more in the past could be the cause of
current QSO activity.

In a twist on the second perspective, \cite{can01} found
spectroscopic evidence for very recent ($\sim$100\,Myr) star formation
in QSOs with a far-infrared excess.
And adaptive optics observations of a few nearby QSOs have also revealed
strong evidence for very recent star formation.
\cite{cre04} argued that the Pa$\alpha$ emission in PG\,1126-041 that
was resolved on spatial scales of 100\,pc was due to star formation.
The inferred star formation rate of 13\,$M_\odot$\,yr$^{-1}$ could
account for most of the object's far-infrared luminosity.
\cite{dav04b} performed a detailed analysis of the unusual galaxy
Mkn\,231 using AO on the Keck telescope.
They showed that the stellar profile and kinematics in the central 800\,pc
were consistent with a rotating disk;
and in this region there was very intense
(50--100\,$M_\odot$\,yr$^{-1}$) star formation which had created
(1--5)$\times10^9$\,$M_\odot$ of stars in the last 10--100\,Myr.
Star formation within 1$^{\prime\prime}$ of the nucleus could account
for about one third of the total luminosity of the galaxy and QSO combined.
In Mkn\,231 it is clear that the star formation has been triggered by
the very recent merger event.
But such a high star formation rate cannot be sustained, and
\cite{dav04b} suggested that once it ceases and fades  -- on timescales
of a few 100\,Myr -- Mkn\,231 will probably look more like a typical
AGN.

\begin{figure}
\begin{center}
\psfig{file=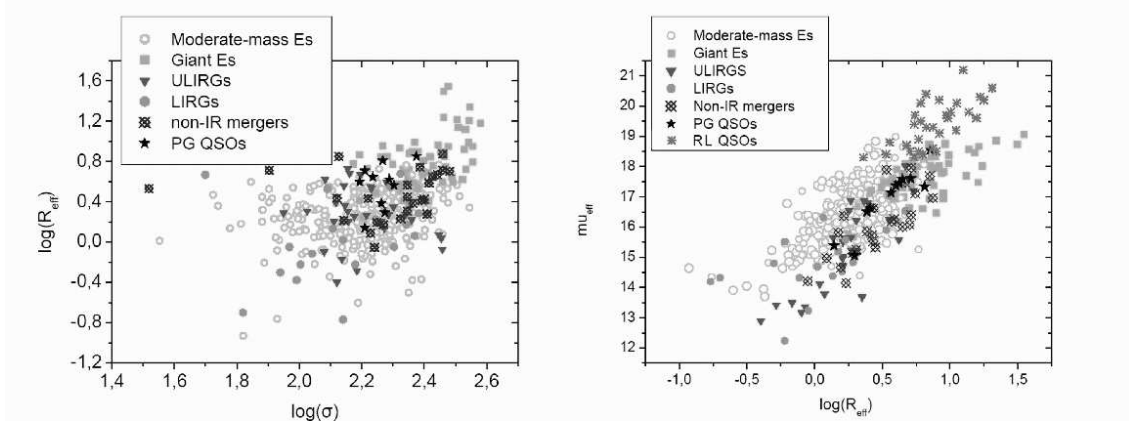,width=15cm}
\end{center}
\caption{Two projections of the fundamental plane for early type
  galaxies, for the effective radius (R$_{\rm eff}$), stellar
  dispersion ($\sigma$), and surface brightness 
(${\rm mu}_{\rm eff}$).
Different symbols and shades denote the location of 
  elliptical galaxies, ULIRGs, and QSOs. 
The IR-bright QSOs (`PG QSOs', black stars) occupy a similar region to
 ULIRGs (red triangles) and cluster ellipticals (`Moderate-mass Es',
 grey circles).
 The QSOs of \cite{dun03} (`RL QSOs', green asterisks) occupy the same
 region as giant ellipticals (`Giant Es', cyan squares). 
Adapted from \cite{das07}, courtesy of K.~Dasyra.}
\label{dav:qso:fig:dasyra}
\end{figure}

Some light can be shed on the issue of QSO hosts by work on stellar
dynamics \citep{das06,das07}.
While this was done without AO, it is certainly something which would
benefit greatly from the technique.
Building on earlier work, these authors measured the stellar
dispersion in a sample of ULIRGs and QSOs, combining the results with
data on the central surface brightness and the host effective radius.
The remarkable result they found (see Fig.~\ref{dav:qso:fig:dasyra})
is not only that ULIRGs lie on the same
fundamental plane as intermediate mass elliptical galaxies, but that
QSOs with infrared excess lie almost in the same region.
This region, however, is significantly displaced from the locus of
radio-loud QSOs, which occupy a position more similar to that of giant
elliptical galaxies.
In addition, the infrared excess QSOs had 
$M_{BH} \sim 10^8$\,$M_\odot$, comparable to those of the ULIRGs, and
hence relatively high Eddington efficiencies of order $\sim$0.25.
These values were considerably different than those inferred for
radio-loud QSOs, which have $M_{BH} \sim
5\times10^8$--$10^9$\,$M_\odot$ and hence Eddington efficiencies
$<0.1$.
Thus, QSOs with infrared excess appear to be intrinsically different
from radio-loud QSOs, suggesting that perhaps there might also be
different causes for their nuclear activity.

Putting all these various results together, one might speculate that there
are in fact two distinct paths to making a QSO.
The first is to merge two gas-rich spiral galaxies, as in the standard
evolutionary scenario.
As has been modelled by many authors, this would lead to massive gas
infall on short timescales, and hence both starburst and AGN activity.
Signs of such events can remain for a very long time, and be observed
as relic evidence of what has occured several Gyr in the past.
However, detecting them does not necessarily mean that the major
merger was directly responsible for all subsequent AGN fuelling.
It is plausible that another event, perhaps accretion of new material
from minor mergers or secular inflow, can also fuel the AGN and lead
to renewed QSO activity several Gyr later.
Certainly, there is still more to learn and understand about what
different processes may lead to QSO activity and the role that the
host galaxy plays in each.
But it is also clear that adaptive optics has the potential to play an
important part in this enterprise.

\section{The Galactic Centre}
\label{dav:gc}

\begin{figure}
\begin{center}
\psfig{file=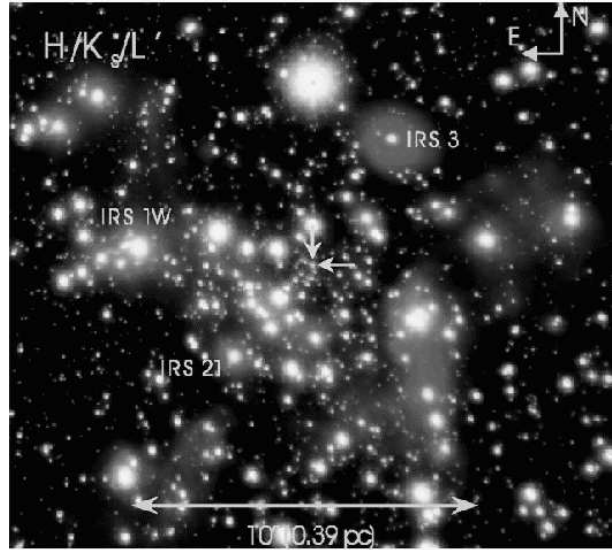,width=8cm}
\end{center}
\caption{Near infrared image of the Galactic Centre, showing the
  detail revealed by adaptive optics in the central half parsec. 
The guide star IRS\,7 for the IR-WFS of NACO is top centre; the
  location of Sgr\,A* is marked.
From \cite{gen03}, courtesy of R.~Genzel.}
\label{dav:gc:fig:gc}
\end{figure}

At the other extreme, in terms of luminosity and distance, we turn to
the Galactic Centre (GC) where there is a black hole of
mass $4\times10^6$\,$M_\odot$ radiating at $<10^{-10}$ of its
Eddington luminosity.
The proximity of this black hole (it is only 8\,kpc distant) means
that even the central parsec around it can be studied in incredible
detail (Fig.~\ref{dav:gc:fig:gc}), making it an obvious way-point on
our quest to learn about AGN. 
Understanding why the massive black hole in the GC is accreting so
little gas, and how this relates to other AGN where we do not have
such a detailed view, must play an important part in piecing together the
whole picture.
As we shall see, adaptive optics has led to rapid developments in our
knowledge of the Galactic Centre and its environment.

\subsection{The Galactic Centre as a Quiescent AGN}
\label{dav:gc:qagn}

At the current time, the GC can be considered a `quiescent AGN'.
The X-ray luminosity centered on Sgr\,A* is
$\sim2\times10^{33}$erg\,s$^{-1}$ within the 
central 0.6$^{\prime\prime}$, implying an accretion rate of
$\sim10^{-6}$\,$M_\odot$\,yr$^{-1}$ on scales of 0.01\,pc
\citep{bag03}.
And measurements of the linear polarisation at radio frequencies
imply an accretion rate near the black hole's event horizon of
$\lesssim10^{-8}$\,$M_\odot$\,yr$^{-1}$ \citep{bow03}.
These are both very much less than the Eddington accretion rate for a
$4\times10^6$\,$M_\odot$ black hole of
$\sim0.01/\eta$\,$M_\odot$\,yr$^{-1}$ where $\eta\sim0.1$ is an
efficiency term.
However, there is also evidence that only a few hundred years ago,
Sgr\,A* was very much brighter.
This comes from detailed X-ray images and spectra which show
3--200\,keV emission from Sgr\,B2, a giant molecular cloud about
100\,pc from Sgr\,A* \citep{koy96,rev04}.
The emission exhibits a 6.4\,keV K$\alpha$ line with a
very large equivalent width, superposed on an absorbed continuum with
strong emission above 20\,keV.
All these characteristics are supported by the interpretation that
hard X-ray emission emitted by Sgr\,A* in the past has been Compton
scattered and reprocessed by the Sgr\,B2 cloud.
The distance between the clouds indicates that the outburst from
Sgr\,A* must have happened 300--400\,yr ago.
And the luminosity of Sgr\,A* at that time would have been
$\sim10^{40}$\,erg\,s$^{-1}$. 
While still only $10^{-5}$\,$L_{Edd}$, this is comparable to low
luminosity AGN which have luminosities in the range
$10^{-3}$--$10^{-6}$\,$L_{Edd}$ \citep{ho99}.
We will re-visit this issue in Section~\ref{dav:gc:faint}, where we
look at why Sgr\,A* is so faint, a property that can be considered an
advantage since it allows us to study the environment close around the
black hole without being dazzled by its glare.

\subsection{Stellar Motions}

At optical wavelengths, the GC cannot be seen as it is hidden behind
$A_V\sim30$\,mag of extinction.
Fortunately, the fact that this corresponds to only $A_K\sim3$\,mag at
near infrared wavelengths, and that there is an incredibly bright
$K\sim6$\,mag star, IRS\,7, less than 6$^{\prime\prime}$ from Sgr\,A*
(see Fig.~\ref{dav:gc:fig:gc})
means that the GC is ideally suited to adaptive optics with an
infrared wavefront sensor.
Over the last decade, adaptive optics -- with visible WFS, infrared
WFS, and LGS -- has driven the huge advances made in the GC, much of
which is related to the stars: their number density, their spectral
types, and their motions.

\begin{figure}
\begin{center}
\psfig{file=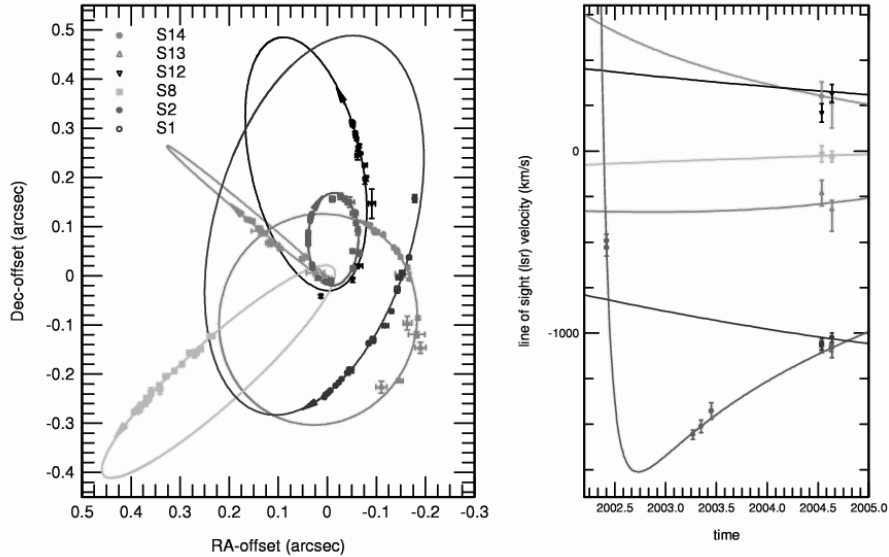,width=12cm,clip=}
\end{center}
\caption{Proper and Radial motions of 6 S stars in the GC, together
  with their orbital solutions. From \cite{eis05}, courtesy of
  F.~Eisenhauer.}
\label{dav:gc:fig:orbits}
\end{figure}

It is tracking of stellar proper motions that is perhaps one of the
most remarkable facets of GC adaptive optics research.
A 15~year (and continuing) campaign has led to measurements
circumscribing a complete orbit of the S2 star, the projected size of
which is only 0.2$^{\prime\prime}$.
Combining the proper motion with radial velocity measurements 
has allowed a fully constrained orbital solution
-- including a geometric distance to the GC -- to be determined
\citep{eis05}. 
These data confirm that the closest approach of S2 to Sgr\,A* was only
17\,light-hours, at which point it had a velocity of
8000\,km\,s$^{-1}$;
and that its orbit has a 15.2\,yr period and a semi-major axis of
0.12$^{\prime\prime}$. 
The latter two measurements yield an estimation of the central mass
that is contained within a few light hours around Sgr\,A* using
Kepler's 3$^{rd}$ law:
\begin{equation}
\left(\frac{P}{2\pi} \right)^2 \ = \ \frac{a^3}{G(M+m)}
\label{dav:gc:eq:kepler}
\end{equation}
It is fascinating to realise that one can still do modern research
using 400-year-old physics -- as long as one has access to
state-of-the-art technology.
This technology has now made it possible to derive useful orbital
solutions for more than 15 stars \citep{ghe05,tri06}, of which 6 are
shown in Fig.~\ref{dav:gc:fig:orbits}.

\subsection{Why is SgrA* so Faint?}
\label{dav:gc:faint}

\begin{figure}
\begin{center}
\psfig{file=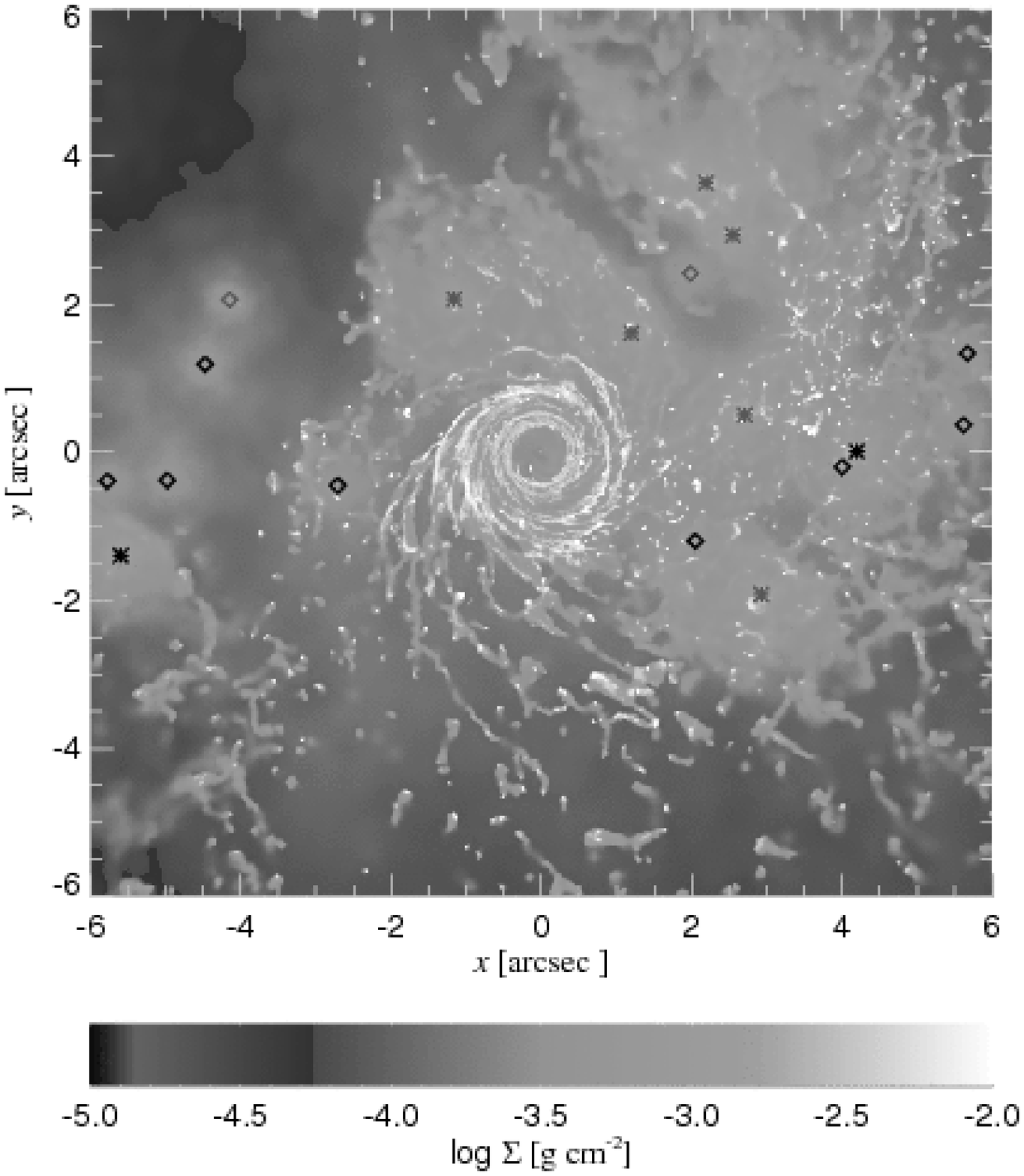,width=6cm}
\psfig{file=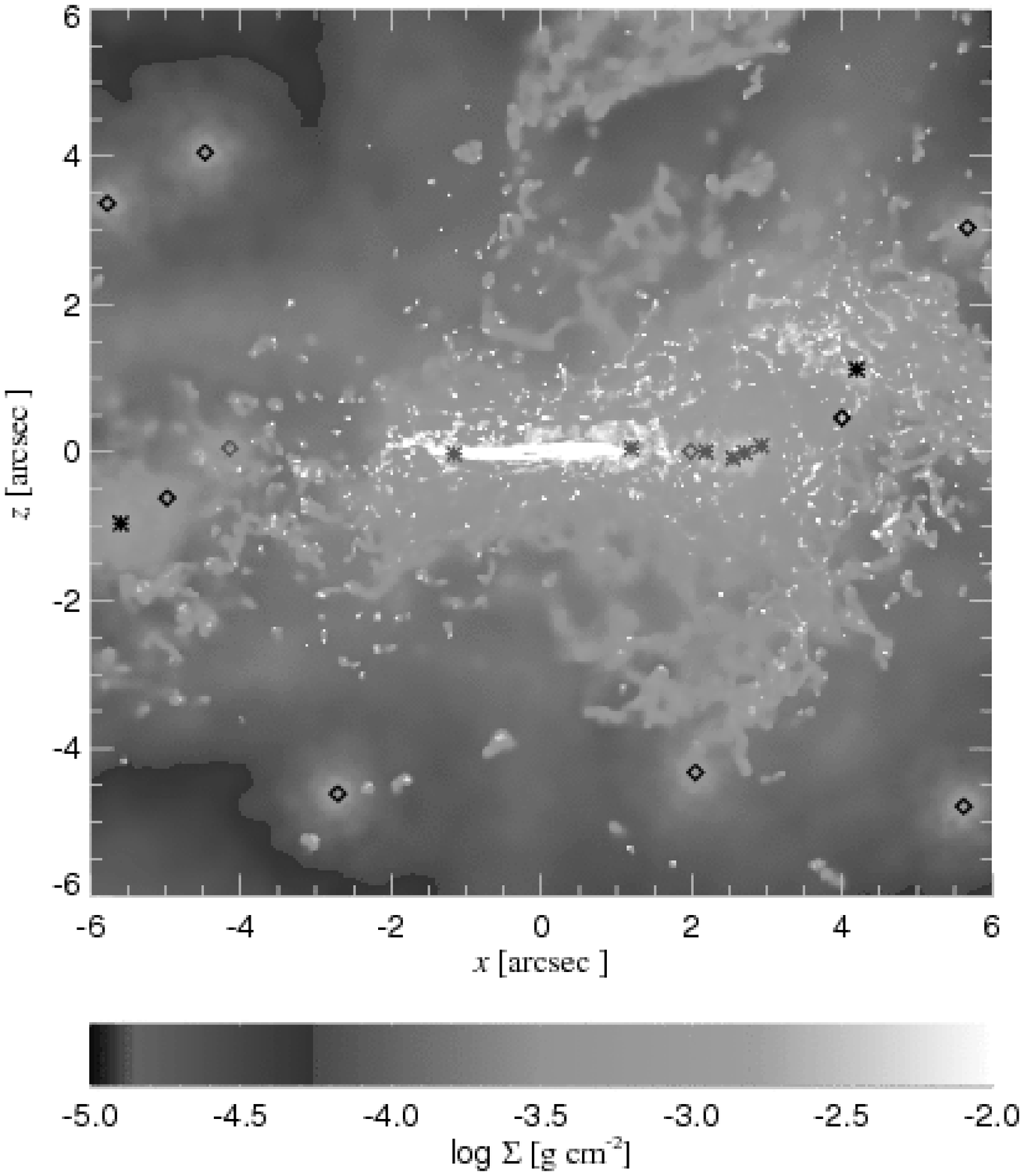,width=6cm}
\end{center}
\caption{End state of the simulation from \cite{cua06} showing the
  column density of gas for 2 projections. 
The positions of the stars are marked by asterisks (for slow winds) or
  diamonds (for fast winds). It can be seen that a hot
  gas phase fills the central cavity and a cold gas phase has settled
  into a disk. Courtesy of J.~Cuadra.}
\label{dav:gc:fig:cuadra}
\end{figure}

It has already been mentioned that Sgr\,A* is remarkably faint.
This ought to be unexpected, since we know that on scales of 10\,pc,
the gas inflow rate is of order $10^{-2}$\,$M_\odot$\,yr$^{-1}$
\citep{mez96}.
This is sufficient material to fuel a Seyfert nucleus.
Yet the actual accretion rate onto the black hole is 6 orders of magnitude
less.
The reason, as proposed by \cite{oze96}, is that inflow is hindered
by the outflow and angular momentum of the winds from massive young
stars that were formed in a recent starburst 5--7\,Myr ago
\citep{pau06}.
This scenario has been modelled in detail by \cite{cua06}.
They included both the fast young stellar winds with velocities of
700\,km\,s$^{-1}$ \citep{oze97} and the slower winds of
$\sim200$\,km\,s$^{-1}$ \citep{pau01}, with a total mass-loss rate of
$\sim10^{-3}$\,$M_\odot$\,yr$^{-1}$.
They also accounted for the 
angular momentum of the stars by putting them on orbits at
$R\sim0.5$\,pc resembling
those that have been observed \citep{gen03,pau06}.
The gas had a two-phase structure.
While much of it flowed outwards, the gas that remained within the
simulation volume consisted of a hot phase filling the
central cavity and a cold phase that condensed into filaments and
fell inwards, settling onto a disk on spatial scales of
about 0.05\,pc (Fig.~\ref{dav:gc:fig:cuadra}).
The simulation yielded two distinct ways in which the black hole could
be fuelled. 
The average accretion rate -- provided by the hot phase -- was only
$\sim3\times10^{-6}$\,$M_\odot$\,yr$^{-1}$, comparable to that observed
on similar scales of 0.01\,pc.
In general, higher accretion rates were hindered by the angular
momentum of the gas.
But superimposed on the base level was a very significant variability
caused by the cold phase as distinct clumps were sporadically
accreted.
More recent simulations \citep{cua08} using stellar properties
specific to individual stars observed in the GC suggest that there are
not enough 
slow winds to form a conspicuous cold disk, but that occasional cold
clumps do nevertheless generate some variability in the accretion rate.
The behaviour in these simulations is reasonably consistent with what
has been observed, and
also has many features in common with models of Seyfert galaxies
which will be discussed in Section~\ref{dav:nearby}.

\subsection{Infrared and X-ray Flares}
\label{dav:gc:flare}

\begin{figure}
\begin{center}
\psfig{file=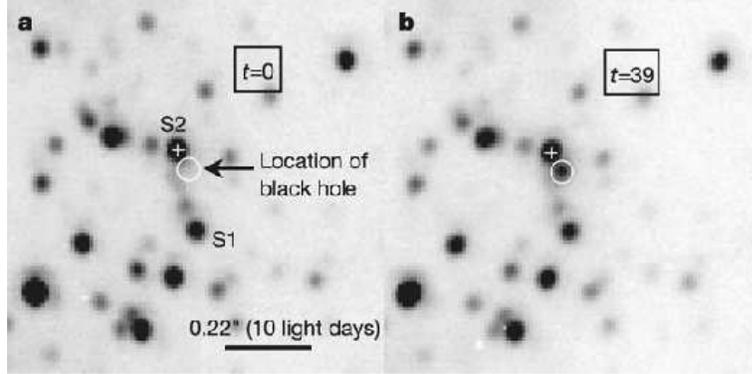,width=10cm}
\end{center}
\caption{H-band images of the Galactic Centre showing the location of
  Sgr\,A* near the centre. A flare from around the black hole has been
  detected in the righthand frame.
From \cite{gen03b}, courtesy of R.~Genzel.}
\label{dav:gc:fig:flare}
\end{figure}

One of the most exciting discoveries in the GC during the last few
years is that of near infrared flares from Sgr\,A* \citep{gen03b}, an
example of which can be seen in Fig.~\ref{dav:gc:fig:flare}.
These occur about twice a day and appear to last for order 1\,hr.
The substructure in the light curve on a time scale of 17\,mins has
been interpreted in terms of matter on the last stable orbit, implying
that the black hole has a spin of $a=0.5$.
In order to understand the physics behind these events, there have
been a number of attempts to obtain spectral information of the
flares.
\cite{gil06} report that the near infrared spectral slope changes with
flare brightness;
\cite{eck06} present coordinated (and sometimes simultaneous)
observations across many wavelength 
regimes (0.7\,cm, 0.89\,mm, 1.7--3.8\,$\mu$m, and 2--8\,keV).
The general consensus of these efforts is an understanding that the
flare is related to, but not directly produced by, the radiatively
inefficient accretion flow (RIAF) that is believed to produce the
peak of the submillimetre emission.
The RIAF model underpredicts the 2\,$\mu$m flux, which instead
originates from electrons that are locally and transiently heated.
As this compact source orbits the black hole, significant variability
ensues from a combination of synchrotron cooling and
orbital dynamics, both of which operate on a timescale of
$\sim20$\,mins.
In this model, the high Lorentz factor of the electrons, $\gamma \sim
10^3$ means that the X-ray emission is most likely to arise from
inverse Compton scattering.

A daring way to observationally constrain such models has been
proposed for GRAVITY, a 4-beam combiner for VLTI that is
currently in its design phase \citep{eis08}.
The two main capabilities of GRAVITY will be phase-referenced imaging
over an 80\,mas field, and narrow angle astrometry to 10\,$\mu$as
accuracy.
The proposal put forward by \cite{pau08} utilises both of these.
The former will provide images of stars very close to Sgr\,A*, and
hence yield proper motions of these stars.
The velocities of these stars are expected to be huge, so that 
within a relatively short time, one should be able to measure orbit
precession due to both the extended mass distribution and also
relativistic effects.
The astrometric capability will enable the centroid of the flares to be
pinpointed to an accuracy corresponding to the Schwarzschild radius
R$_S$ with less than 1\,min integration time.
Thus it will be possible to track the motion of the flare centroid as
it evolves; and, if each flare traces the same orbit, co-add
observations from different flares to increase the signal-to-noise.
Relativistic ray tracing indicates that if the flare does indeed
originate from a compact spot of transiently heated electrons, then
such methods will be able to constrain the geometry and physical
conditions prevailing at distance of a few R$_S$ around the Galactic
Centre's black hole.

\section{Nearby Active Galactic Nuclei}
\label{dav:nearby}

Nearby galaxies -- at distances of a few to a few tens of Mpc -- present an
opportunity to observe the AGN phenomenon in a larger number of objects on
scales suited to studying some of their major structures.
These AGN are almost exclusively Seyfert galaxies, which are less
luminous then QSOs but are believed to fit into the same unification
scheme.
In such galaxies, AO allows us to reach spatial resolutions
corresponding to 1--50\,pc.
One is therefore able to resolve the circumnuclear region of the host
galaxy, the Narrow Line Region, the obscuring molecular torus and the
nuclear region of the host galaxy, and in some cases the sphere of
influence of the central supermassive black hole.

The Seyfert galaxy NGC\,1068 at a distance of 14\,Mpc is perhaps an
obvious choice, not only due to its proximity, but because it is
considered to be `prototypical' and has one of the brightest nuclei.
It has been the subject of many studies, even from the early
days of adaptive optics 
\citep{rou98, rou04, mar00, gra03, gra05, gra06, mue08} and interferometry 
\citep{jaf04}.
A few Seyfert galaxies lie nearer still. 
In particular, at distances of only $\sim$4\,Mpc -- where 60\,mas, the
K-band diffraction limit of an 8-m telescope, corresponds to 1\,pc --
are the Circinus Galaxy and Centaurus\,A.
Wavefront sensing on these objects is more challenging, and it has
only been possible relatively recently to observe and study them with
AO \citep{pri04, mue06, neu07} or interferometry
\citep{tri07, mei07}.
Adaptive optics systems have also been pointed towards a few other
AGN, including 
Mkn\,231 \citep{lai98, dav04b},
M\,81 \citep{dav99},
NGC\,6240 \citep{bog03, max05, pol07},
NGC\,7469 \citep{dav04a}, 
NGC\,1097 \citep{pri05}, and 
NGC\,3227 \citep{dav:dav06}.

A few of these objects were also included in a sample of nearby AGN
observed by \cite{hic08a} to study molecular gas dynamics with an
aim to estimating their black hole masses.
A number were also included in the work that makes up much of what is
presented in this Section, based on 
observations with the adaptive optics near infrared integral field
spectrometer SINFONI \citep{eis03,bon04}.
This sample comprises 9 galaxies which were observed using the AGN
itself as the AO wavefront reference \citep{dav:dav07,dav:hic08b}.
The open questions that we will address with the data here aim to
(1) measure the black hole mass using spatially resolved stellar
kinematics;
(2) derive the properties of the molecular gas, and understand
its relation to the torus; and
(3) determine the extent and history of star formation, and its
relation to the AGN and torus.

\subsection{Black Hole Masses}
\label{dav:nearby:bh}

There are many ways to measure black hole masses depending
on how far away the AGN is and hence what spatial scales can be
resolved.
For nearby AGN, stellar or gas kinematics represent a good option.
Gas kinematics are often easier to model since gas cools into a thin
rotating disk on short timescales, and can usually be modelled as
such.
However, gas can also be strongly influenced by inflows and outflows.
For that reason, stellar kinematics are preferred, although the
modelling is more complex.
In either case, an important issue is that the radius of influence of
the black hole $R_h$ is resolved.
This is defined as the radius within which the black hole mass, rather than
the stellar mass, dominates the gravitational potential, and is
\citep{fer05}
\begin{equation}
R_h \ \sim \ \frac{GM_{BH}}{\sigma_*^2} 
    \ \sim \ 11.2 \left( \frac{M_{BH}}{10^8\,M_\odot} \right)
                  \left( \frac{\sigma_*}{200\,km\,s^{-1}} \right)^{-2}
    \ pc
\label{dav:nearby:eq:r_bh}
\end{equation}
where $\sigma_*$ is the velocity dispersion of the stellar spheroid
around the black hole.
On the other hand, we know from the $M_{BH}$-$\sigma_*$ relation
that \citep{tre02}
\begin{equation}
M_{BH} \ = 1.35\times10^8 
       \left( \frac{\sigma_*}{200\,km\,s^{-1}} \right)^{4.02} \ M_\odot
\label{dav:nearby:eq:msig}
\end{equation}
And putting Eqs.~\ref{dav:nearby:eq:r_bh} and~\ref{dav:nearby:eq:msig}
together one gets, to zero order, an estimate of $R_h$ which depends
only on $M_{BH}$
\begin{equation}
R_h \ \sim \ \left( \frac{M_{BH}}{10^6\,M_\odot} \right)^{0.5} \ pc
\label{dav:nearby:eq:rh}
\end{equation}
Thus, with adaptive optics on an 8-m telescope, one can resolve $R_h$ in
a typical Seyfert galaxy harbouring a $10^7$\,M$_\odot$ black
hole out to a distance of $\sim20$\,Mpc.

Many black hole mass estimates have been derived from the
line-of-sight velocity distributions (LOSVDs) extracted from long-slit
data along particular axes.
In contrast, integral field data maps the complete spatial distribution of
the LOSVDs, and so is more robust.
The reason, explained clearly by \cite{cap04}, is that long-slit data
may miss some orbits if the (localised) regions of high projected
surface brightness do not happen to fall within the slit.
As a result, a model derived using long-slit data may be
significantly different from that found from integral field data.

The LOSVD itself is usually characterised in terms of Gauss-Hermite
functions.
These include the velocity $V$ and dispersion $\sigma$ of a simple
Gaussian, as well as
the 3$^{rd}$ and 4$^{th}$ Hermite terms which quantify
respectively the asymmetry and flatness of the profile
\citep[see][for a more complete description of these]{ben94}.
The LOSVDs are extracted by deconvolving the galaxy spectrum using
spectra of local stellar templates.
Typically several are employed to reduce problems of template
mismatch due to the varying depths of the different absorption
features.
There are two commonly used methods: division of the galaxy spectrum
by the template in the Fourier domain; 
or direct convolution of a function with the template to match the
galaxy spectrum.
Now that computer power is not a limiting issue, the latter method is
increasingly used, and it leads to either a full non-parametric
LOSVD or to a parameterisation at the desired level of detail.
A discussion of this process, including a detailed analysis of one
particular method, is given by \cite{cap04b}.
Finally, once the LOSVDs have been extracted, one can begin to model them.

The kinematics of stellar spheroids are modelled using an orbit
superposition method developed originally by \cite{sch79}, and 
extended by \cite{cre99} to the more general three
integral axisymmetric models that are commonly used now.
At the start of each step one specifies the black hole mass and the
mass-to-light ratio 
of the stars, and uses those together with the observed luminosity
distribution to define the gravitational potential.
One then follows the equations of motion to build up a library of
several thousand pro- and retro-grade orbits distributed through
parameter space (comprising energy, angular momentum, and a more
abstruse `third integral').
A linear non-negative combination of orbits is then found that
matches the luminosity profile and reproduces as closely as
possible the observed LOSVDs.
One can then repeat the process using different input parameters:
black hole mass, mass-to-light ratio, inclination, and so on.
There are inevitably details which need to be carefully considered
\citep[see, for example,][and references therein]{tho04}, but the
final result is that, via a $\chi^2$ minimisation, one can find the
optimal parameters as well as their associated uncertainties.

\begin{figure}
\begin{center}
\psfig{file=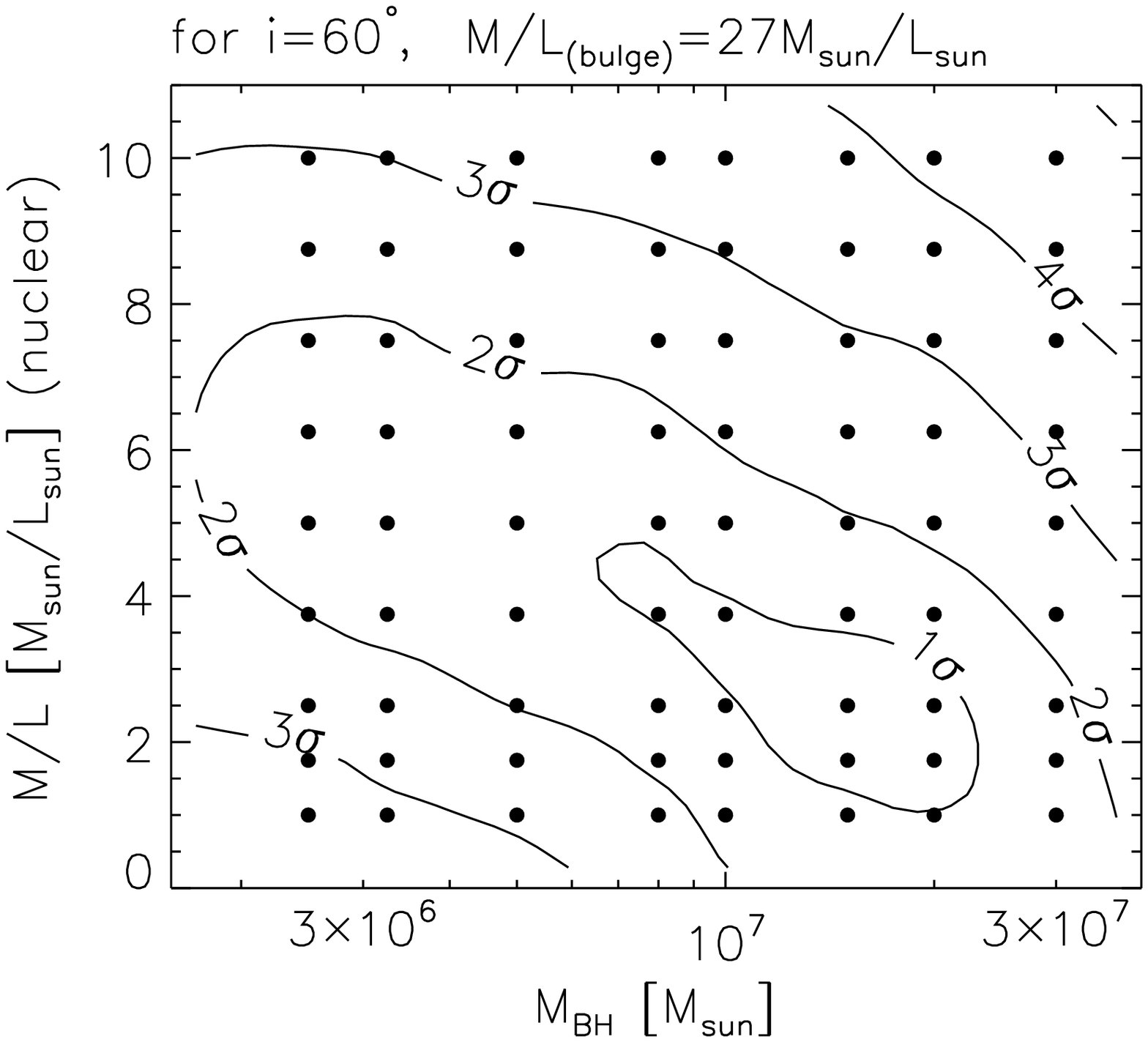,width=4.5cm}
\psfig{file=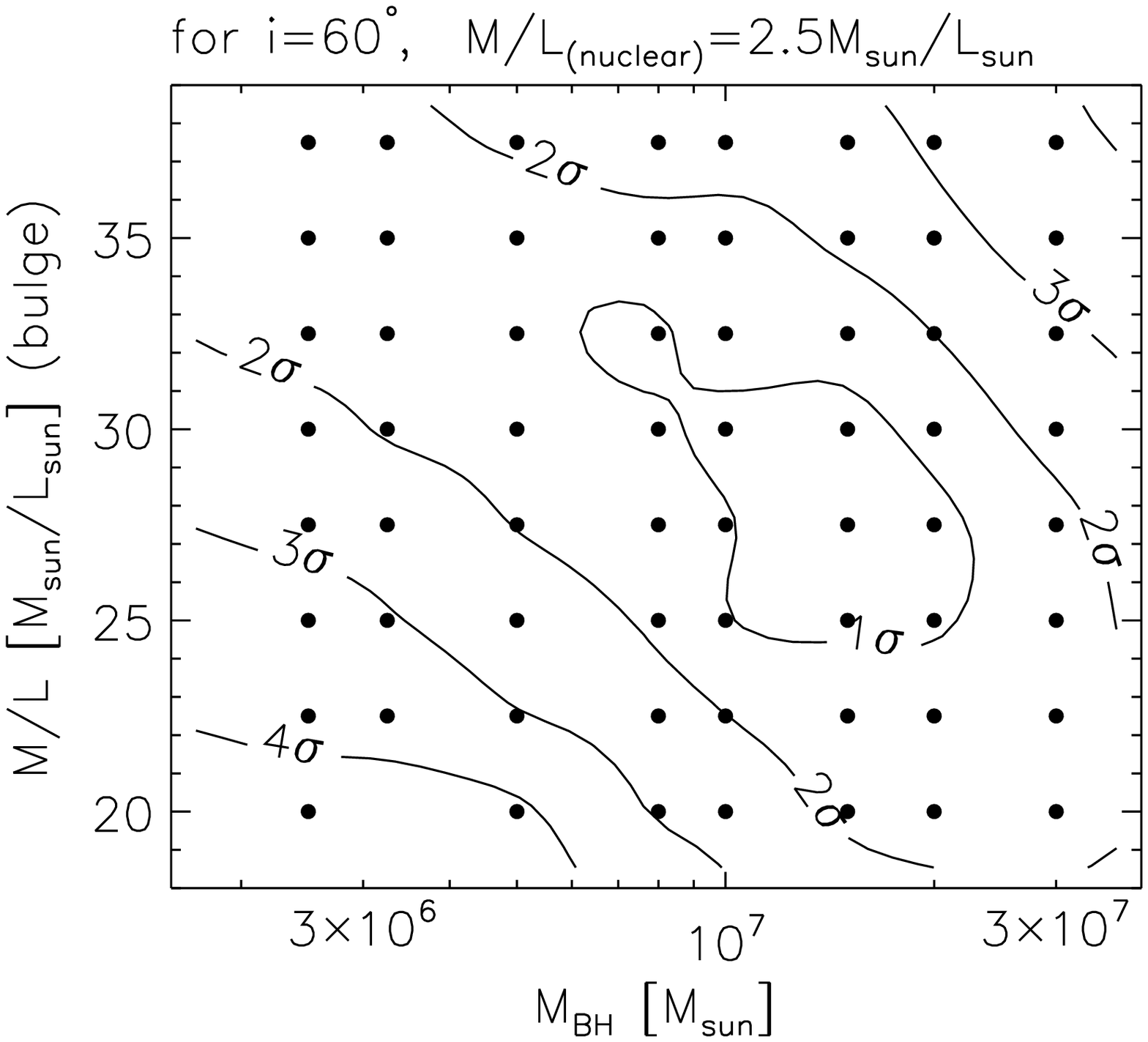,width=4.5cm}
\psfig{file=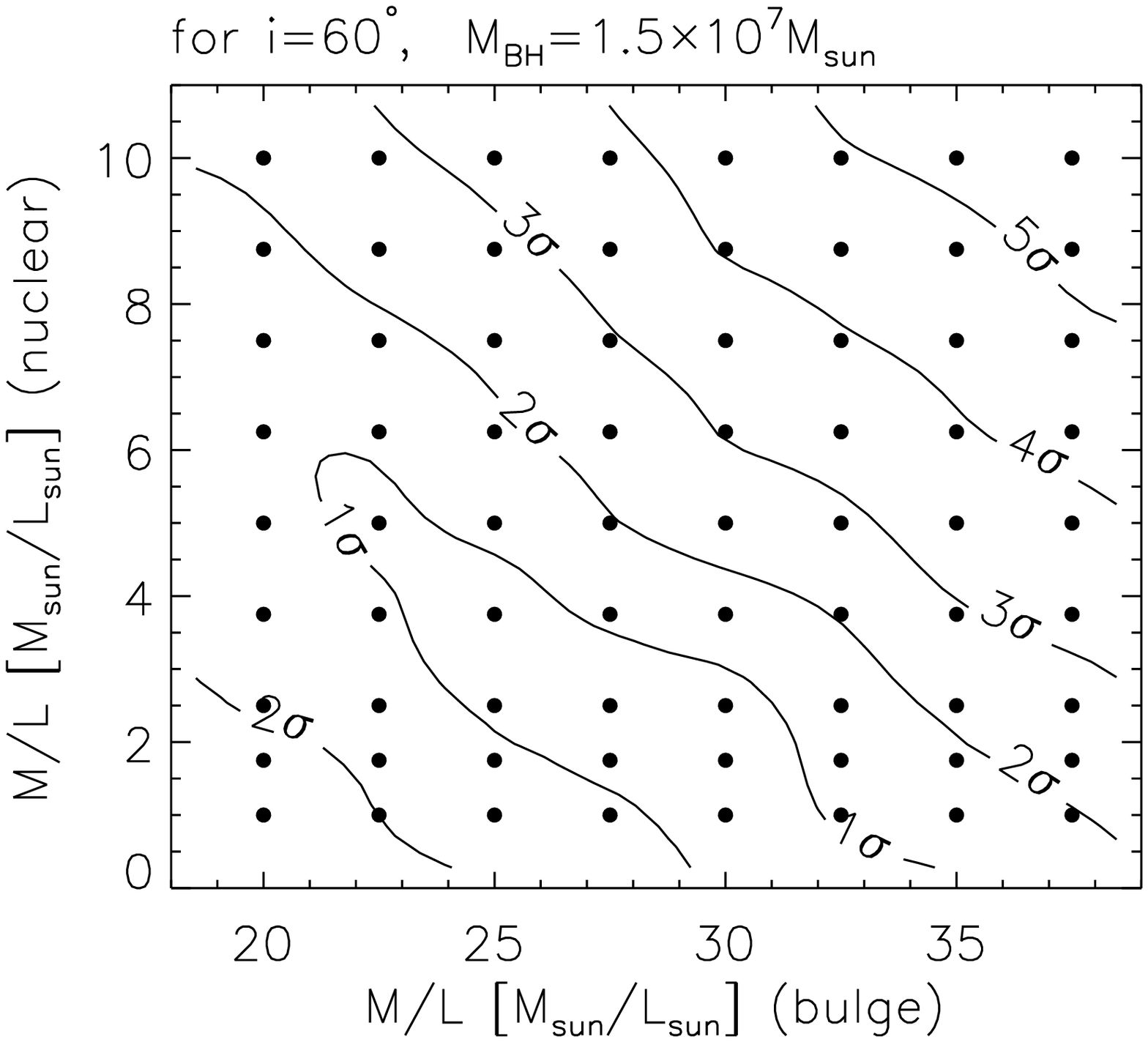,width=4.5cm}
\end{center}
\caption{Slices through the $\chi^2$ landscape for the Schwarzschild
  modelling performed on NGC\,3227, showing how the 3 parameters are
  correlated. 
Left: at a fixed $M/L_{bulge}=27$\,$M_\odot/L_\odot$.
Centre: at a fixed $M/L_{nuclear}=2.5$\,$M_\odot/L_\odot$.
Right: at a fixed $M_{BH}=1.5\times1-^7$\,$M_\odot$.
From \cite{dav:dav06}.}
\label{dav:nearby:fig:n3227}
\end{figure}

The example of NGC\,3227 \citep{dav:dav06} demonstrates several
important issues that should be borne in mind.
For this object, two stellar components were included, representing
the bulge of older stars and a thick nuclear disk of younger stars.
The resulting $\chi^2$ landscapes for the three parameters ($M_{BH}$,
$M/L_{bulge}$, and $M/L_{nuclear}$) is shown in
Fig.~\ref{dav:nearby:fig:n3227}.
Firstly, each pair of parameters are apparently correlated.
This is because the mass on small spatial scales can either be in a
black hole or in the nuclear disk; and on slightly larger scales, the
mass can be distributed either in the disk or bulge.
Thus, the parameters are not entirely independent.
Secondly, it suggests that the derived black hole mass
should instead be considered as the unresolved central mass, which may
include a compact stellar component. 
Some work along these lines has been pursued by \cite{fer06}, but the
topic remains highly controversial.
Thirdly, consistency checks are important.
The NGC\,3227 model yielded $M/L_{bulge} = 27.5$\,$M_\odot/L_\odot$,
within the range 25--35\,$M_\odot/L_\odot$ expected for bulges of
spiral galaxies \citep{for03}.
On the other hand, it suggested that 
$M/L_{nuclear} = 2.5$\,$M_\odot/L_\odot$, a factor of 5 larger than that
derived from modelling the starburst using evolutionary synthesis
code.
The reason is that the mass of the thick nuclear disk
includes a substantial contribution from molecular gas, so that its
effective ratio is $M/L_{nuclear} = 1$--5\,$M_\odot/L_\odot$.
Thus the model result is consistent with expectations.
It is such consistency checks that can give one confidence that the
derived black hole mass -- in this case $1.5\times10^7$\,M$_\odot$ --
is reliable.

\subsection{Obscuring Molecular Gas}
\label{dav:nearby:torus}

To understand how the molecular gas seen in the SINFONI data is
related to the obscuring molecular torus, one first needs to clear
one's mind of any preconcieved ideas about how the torus should look.
Instead, we must think only of the minimum global
criteria that are necessary for a structure to be described as the
torus.
These are that it
(1) consists of molecular gas and dust;
(2) is compact, with a size no more than several tens of parsecs;
(3) is optically thick, so that it obscures the AGN when edge-on;
(4) is vertically extended, by several parsecs, so as to provide
collimation for ionisation cones.
As we summarise briefly below, \cite{dav:hic08b} argues that these
criteria are indeed fulfilled.

The near infrared molecular gas tracer is the 1-0\,S(1)
ro-vibrational line of H$_2$, and
arises in gas that is heated to temperatures of (1--2)$\times10^3$\,K.
Since this can only occur at the edges of clouds, the line traces only
$\sim10^{-6}$ of the total gas mass, and on small scales can be
strongly influenced by the local environment (e.g. proximity of young
stars, supernova remnants, etc.).
Nevertheless, on large scales it is believed to trace reasonably well
the distribution of molecular gas -- and its presence indicates
that there is certainly molecular gas there, as one would expect for
the torus.
The size scale can be found from the half-width at half maximum of the
radially averaged profiles of the line.
These are typically less than 50\,pc, and in many cases much less.
The optical depth, or equivalently column density, is harder to derive
because the line traces such a small fraction of the total gas mass.
However, it can be estimated from the dynamical mass (based on the
observed velocity and dispersion), assuming an
appropriate gas mass fraction.
\cite{dav:hic08b} argue that 10\% is a conservative fraction, and even this
implies a column density of more than $10^{23}$\,cm$^{-2}$.
Given that extinction $A_V$ is related to the gas column density $n_H$
as $A_V \sim n_H ({\rm cm}^{-2}) / 10^{21}$, this is more than enough
to obscure the AGN even at near infrared wavelengths.
Evidence for a vertically extended geometry comes from the large
50--100\,km\,s$^{-1}$ velocity dispersion of the gas.
The 1-0\,S(1) line emission is strongest if it is excited in shocks
with speeds of 20--40\,km\,s\,$^{-1}$.
Higher velocities would dissociate the H$_2$ molecules and reduce the
1-0\,S(1) line intensity.
However, high dispersions can be produced if there are bulk motions --
such as dense clouds moving at high speeds, as observed in
Orion \citep{ted99}.
There the 1-0\,S(1) emission originates in oblique shocks on the sides of
the clouds, and the width of the line, when integrated over the cloud,
yields a measure of the speed of the cloud.
A very broad 1-0\,S(1) line is therefore an indication that there are
oblique shocks occurring on many clouds which are moving in different
directions, suggesting that the  molecular gas seen close around AGN
is vertically extended.

Thus, it is fair to conclude that adaptive optics observations of
nearby AGN do resolve the global structure of the torus.
In Section~\ref{dav:nearby:sftor} we shall see how these large
scale properties can be reconciled with the small scale structures
inferred from interferometric observations.

\subsection{Nuclear Star Formation}
\label{dav:nearby:sf}

Over the last decade, there has been increasing evidence that
starbursts do occur close to a majority of AGN.
This has led to a framework in which the fundamental questions concern
the extent to which AGN and nuclear star formation might impact
each other via their feedback mechanisms.
Since this is closely related to how 
energetically significant starbursts are around AGN,
considerable effort has been devoted to studying
starbursts in local Seyfert nuclei.
The largest of these studies \citep{cid04} made use of empirical
population synthesis, finding
unambiguous evidence for recent ($<$1Gyr) star formation within a few
hundred parsecs of the nuclei in about 40\% of Seyfert~2 nuclei.
However, to address whether there might a causal relationship, one
needs to probe the star formation in more detail.
This is where adaptive optics plays a crucial role, not only providing
higher spatial resolution, but doing so in Seyfert~1 nuclei where
historically the glare from the AGN itself has overwhelmed any
starburst signature.

Such work with AO was the focus of a study by \cite{dav:dav07} on the
sample of AGN observed with SINFONI.
The authors first explain how to separate the
stellar and non-stellar contributions to the near-infrared continuum
using the equivalent width of stellar absorption features.
They go on to show that one can in principle estimate the stellar
bolometric luminosity to within a factor of 3 without knowing anything
about the star formation history, although
they do use several independent diagnostics to better
constrain the star formation age.
Their findings are that 
(1) the nuclear star forming regions are spatially resolved, with
half-widths of 50\,pc or less;
(2) there has been recent star formation within the last few hundred
Myr;
(3) the star formation is no longer active, implying that nuclear
starbursts are short-lived phenomena.
The way in which this star formation might be linked to both the torus
and to the AGN itself is the topic of the following two sections.

\subsection{The Star-Forming Torus}
\label{dav:nearby:sftor}

\begin{figure}
\begin{center}
\psfig{file=fig13a.eps,width=11cm}
\psfig{file=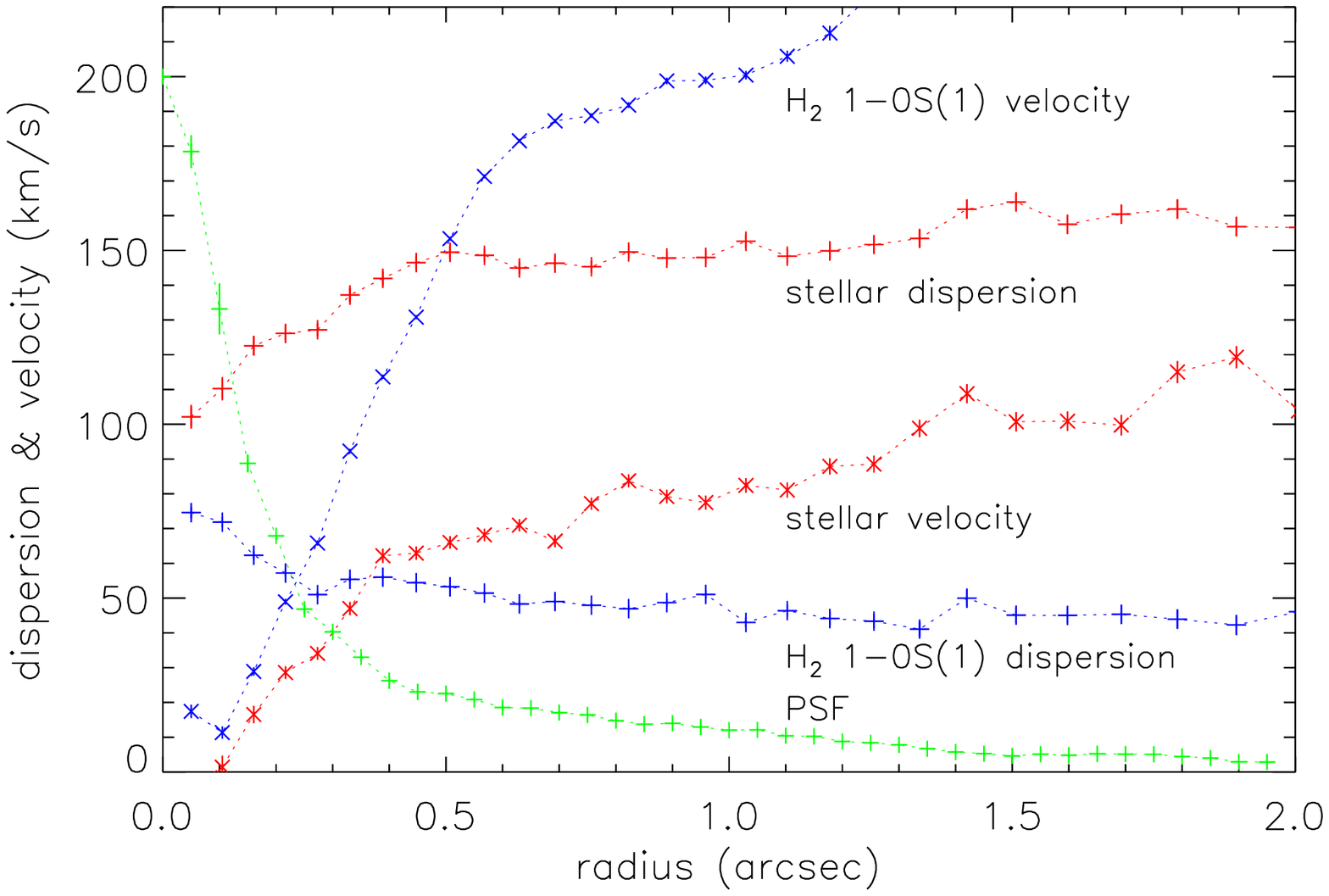,width=10cm}
\end{center}
\caption{Distribution (top) and azimuthally averaged kinematics
  (bottom) for the stars and molecular gas in NGC\,1097. 
Axes are marked in parsecs (top) and arcsec (bottom). 
For NGC\,1097, 1$^{\prime\prime}$ corresponds to 80\,pc.
Adapted from \cite{dav:dav07}.}
\label{dav:nearby:fig:n1097}
\end{figure}

In the previous two sections, we have seen that both the gas and stars
in the nucleus have half-widths less than about 50\,pc.
Although not conclusive in itself, this suggests that the stars and
gas (and hence the torus) may be related.
However, it is the kinematics that are needed to decide whether they
actually occupy the same volume. 
NGC\,1097 (Fig.~\ref{dav:nearby:fig:n1097}) is a clear example where
this appears to be the case.
The rotation curves of the gas and stars are completely different at
radii $>0.5^{\prime\prime}$.
When interpreted with the dispersion, they indicate that the stars are
in a spheroidal structure dominated by random motions (i.e. the bulge)
whereas the gas is in a thin disk supported by ordered rotation.
But at radii $<0.5^{\prime\prime}$, the gas rotation curve drops
rapidly and quickly becomes very similar to that of the stars.
Similarly, the dispersion of the stars drops while that of the gas
increases.
Thus the stellar and gas kinematics become remarkably similar at
small radii.
The only possible explanation is that in the nucleus, the gas and stars
are physically mixed.

If one draws all the various threads together, one is led to the intriguing
conclusion that starbursts close around AGN are intense and
short-lived events that recur episodically.
The hypothesis of \cite{dav:dav06,dav:dav07}
is that gas accumulates in the central 100\,pc region, 
either gradually or by stochastic accretion onto these scales of
a number of giant molecular clouds.
At first stars cannot form, because the interstellar medium is too
turbulent; but eventually as the gas density increases, star formation
does ensue.
This is simply a statement of the Toomre criterion
\begin{equation}
Q \ = \ \frac{\sigma \kappa}{\pi G \Sigma}
\label{dav:nearby:eq:toomre}
\end{equation}
where the relevant parameters here are the dispersion $\sigma$ and
mass surface density $\Sigma$.
The value of $Q$ is a balance between high mass density facilitating
star formation, and turbulence hindering it.
Only when $Q<1$ can star formation take place.
The global Schmidt law formulated by \cite{ken98} states
\begin{equation}
\Sigma_{SFR} \ \propto \ \Sigma_{gas}^{1.4}
\label{dav:nearby:eq:ken98}
\end{equation}
so that when star formation does occur, the gas mass density is so high that
star formation is extremely rapid and efficient.
This will mean that the starburst is inevitably short-lived.
And the combined effects of radiation pressure, and later
turbulence induced by supernovae, will also help to shut off the star
formation.
The starburst will fade relatively quickly, and the region will then
remain dormant until the gas is replenished by further inflow to the
central 100\,pc.
While this scenario may sound plausible, it still needs to be verified
theoretically, and much work also remains on understanding in detail
how such a scheme might work.

\subsection{The AGN-Starburst Connection}
\label{dav:nearby:agnsf}

One consequence of the $M_{BH}-\sigma_*$ relation, that black holes and
bulges grow in tandem, is that understanding how feedback moderates
star formation and black hole growth has become an increasingly
relevant issue.
We have already enountered this in Section~\ref{dav:qso:youngstars},
and it appears in many other arenas.
One of these concerns how early type galaxies evolve from the `blue
cloud' to the `red sequence' -- regions on plots of the stellar mass
against colour.
This evolutionary path has been studied by many authors, notably
\cite{sch07}.
These authors took a sample of 16000 early type galaxies from the
Sloan Digitised Sky Survey, and measured their AGN properties (via
emission lines) and star formation history (via absorption features
and multi-colour photometry).
They could very clearly separate the star forming galaxies which
inhabited the `blue cloud' from the quiescent galaxies which occupied
the `red sequence'.
But they also found evidence of an evolutionary sequence in which the
Seyfert galaxies occupying the `green valley' in between 
exhibited characteristics of intermediate age star formation.
Their analysis showed that the peak AGN phase occurs roughly 0.5\,Gyr after
the starburst, and they suggested that this was due to suppression of
the star formation (on large scales) by the AGN feedback.

\begin{figure}
\begin{center}
\psfig{file=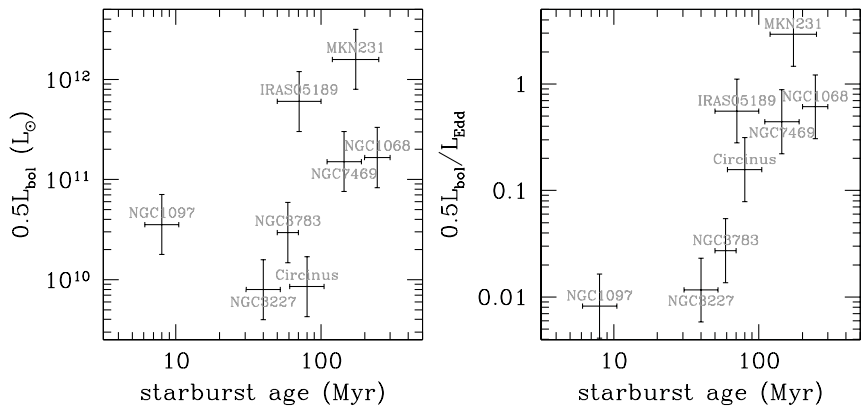,width=12cm}
\end{center}
\caption{Graphs showing how the luminosity of an AGN (left) or its
  accretion rate (right) might be related to the age of the most
  recent episode of star formation.
From \cite{dav:dav07}.}
\label{dav:nearby:fig:delay}
\end{figure}

A rather different conclusion -- that the starburst (on very small
scales) has a considerable
impact on the AGN fuelling -- was reached by \cite{dav:dav07} from the
AO data discussed in Section~\ref{dav:nearby:sf}.
This analysis produced a qualitatively similar result, shown in
Fig.~\ref{dav:nearby:fig:delay}, that the AGN appears to be switched on
50--100\,Myr after the starburst has begun.
However, it was interpreted as a delay of the AGN fuelling induced
by the starburst.
A crucial aspect of this may be the stellar ejecta; and in particular
not just the mass loss rate but the ejection speed.
In Section~\ref{dav:gc:faint} we saw that in the Galactic Centre,
winds from OB stars are hindering gas accretion onto the black hole.
Similarly, it is likely that supernovae, each of which may eject
5\,$M_\odot$ of mass at 5000\,km\,s$^{-1}$, will also do so.
It is known that starbursts often lead to superwinds in which a
significant fraction of the interstellar medium is ejected -- perhaps
the best known example is M\,82.
For a starburst of short duration, the supernova rate peaks at
ages of 10--50\,Myr.
Fig.~\ref{dav:nearby:fig:delay} shows that this is the period when the
AGN are in a low luminosity phase, supporting the idea that supernovae
hinder accretion.
On the other hand, stars of 1--8\,M$_\odot$ reach the end of their
main sequence lives after $\sim50$\,Myr, and evolve into thermally
pulsing asymptotic giant branch (AGB) stars.
Their winds, which have much slower speeds of
10--30\,km\,s$^{-1}$, will remain bound.
And because they are still associated with significant mass loss,
could, in principle, provide sufficient fuel to power a Seyfert
nucleus.

\begin{figure}
\begin{center}
\psfig{file=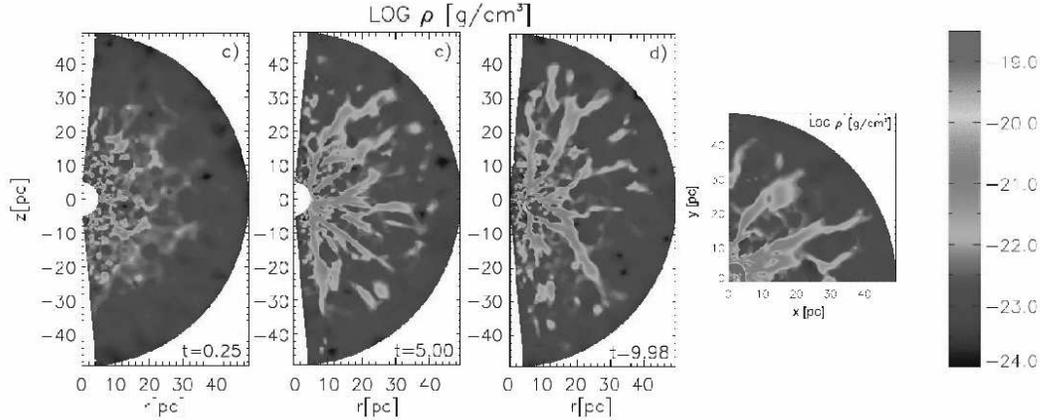,width=14cm}
\end{center}
\caption{Hydrodynamical simulations of the impact of stellar evolution
  on the gas around an AGN. Left three panels: gas density in
  meridional plane at the start and after 5 and 10 orbits.
Right panel: gas density in the equatorial plane after 10 orbits.
The filamentary structure and compact turbulent disk are clearly seen.
Adapted from \cite{msch07}, courtesy of M.~Schartmann.}
\label{dav:nearby:fig:sch}
\end{figure}

Intriguingly, hydrodynamical simulations of the impact of stellar
evolution on gas in the central 50\,pc of AGN have led \cite{msch07}
to a remarkably similar conclusion to that above.
He found that supernovae can blow low density gas away, leaving
long dense filaments (Fig.~\ref{dav:nearby:fig:sch}).
The interplay between these filaments and the slower stellar ejecta
lead to inward accretion to form a central turbulent disk.
There are 3 insights that one can immediately draw from this work.
The first is that a starburst does indeed have a dramatic impact on
the nuclear region, and that gas ejected in slow winds can be accreted
along filaments.
Secondly, there is a hint that the torus may actually have
substructure: a larger scale diffuse part consisting of long filaments,
and a small scale dense turbulent disk.
It would be the former structure that the AO data have found; and the
latter that shows up most prominently in interferometric observations
of AGN.
Finally, the processes operating in this simulation of AGN are
basically the same as those seen in the Galactic Centre
(Fig.~\ref{dav:gc:fig:cuadra}).
The main difference is that of spatial scale which may, as suggested
by \cite{tho05}, be related to the mass of the central black hole.
With hindsight, the similarities should perhaps not be surprising,
since the physics in both environments concerns the impact of stellar
winds on the ISM and inward accretion.
It is a salutary lesson that there is much to be learnt by
adopting a broad perspective.

\section{Outlook}
\label{dav:conc}

In these lectures we have taken a brief look at how adaptive optics
works, and how it has led to new insights in studies of AGN on many
spatial scales.
At high redshift, AO enables one to resolve the host galaxy of AGN on
kpc scales.
In the Galactic Centre, it opens up a detailed view on the sub-parsec
scale.
And in nearby galaxies, it yields a resolution comparable to the size
scale of a number of the structures which make up an AGN.

The next few years will be a very exciting time, as new advanced AO
systems become available to the community on several different
large telescopes.
These include the multi-conjugate laser guide star system on
Gemini South, which will use 5 LGS and 3 NGS to control 3 DMs, and
provide diffraction limited resolution at 1-2.5\,$\mu$m across a field
of view of more than 1\,arcmin.
LINC-NIRVANA on the Large Binocular Telescope will use multiple NGS to
flatten the wavefront so that the beams from the two mirrors can be combined
interferometrically, yielding images with a resolution equivalent to
that of a 23-m telescope.
MUSE on the Very Large Telescope will make use of the multiple laser
guide stars available with the Adaptive Optics
Facility to provide either enhanced resolution over 1\,arcmin, or
diffraction limited resolution over a smaller field -- both at optical
wavelengths.
These systems will open up many possibilities for studying AGN across
all spatial scales.
The progress that can be made will be limited only by the imagination
of those who wish to use them.

\section*{Acknowledgements}

It is a pleasure to thank all those who have generously contributed
figures, and to those who have patiently commented on the text. 
I would also like to thank the referee for a number of helpful suggestions. 
Finally, I thank the school organisers for inviting me to give these
lectures.


\end{document}